\documentstyle[aps,preprint,eqsecnum]{revtex}
\begin{document}
  \draft
\title{The U(1) Gross-Neveu Model at Non-Zero Chemical Potential}
\author{Simon Hands}
\address{Department of Physics, University of Wales, Swansea,\\
Singleton Park, Swansea, SA2 8PP, U.K.}
\author{Seyong Kim}
\address{School of Physics,
High Energy Physics Division\\
Argonne National Laboratory\\
9700 S. Cass Avenue,
Argonne, IL 60439}
\author{John B. Kogut}
\address{Department of Physics,
University of Illinois at Urbana-Champaign\\
1110 West Green Street,
Urbana, IL 61801-3080}
\maketitle

\begin{abstract}
The four-fermi model with continuous chiral
symmetry is studied in three dimensions at non-zero chemical potential $\mu$
using both the $1/N_f$ expansion and computer simulations.
For strong coupling this model
spontaneously
breaks its U(1) chiral symmetry at zero chemical potential and the Goldstone
mechanism is realized through massless pions.  The computer simulation
predicts a critical chemical potential $\mu_c$ close to the
lightest fermion mass in the model.  As $\mu$ is increased beyond $\mu_c$, the
pion screening mass increases rapidly from zero to a nonvanishing value
indicating symmetry restoration.
Some lessons are drawn relevant to lattice QCD
simulations at non-zero $\mu$.
\end{abstract}
\vfill
\pagebreak

\section{Introduction}
Simulations of lattice QCD at non-zero chemical potential $\mu$
have remained in
a quagmire for over a decade~\cite{Barb1}.
The lattice action becomes complex when
$\mu$ is taken
non-zero, so conventional computer simulation algorithms which are based on
a probability distribution
do not apply.  The quenched version of lattice QCD, which
attempts to
skirt the issue by replacing the complex fermion determinant by unity,
appears to be
pathological in the chiral limit~\cite{Barb2}\cite{DK}.
When the bare quark mass is non-zero so
the pion is
not massless, one finds that quenched simulations of lattice QCD work for $\mu$
outside a ``forbidden" region extending from $m_\pi /2$ to $m_B /3$,
where $m_{\pi,B}$ denote pion and baryon masses respectively~\cite{KLS}.  The
failure of
the quenched version of QCD to describe the forbidden region is poorly
understood~\cite{Vink}\cite{BDP}\cite{Mendel}.
Studies of zero dimensional models of non-zero chemical potential have not been
decisive -- some models work in their quenched versions \cite{BD}
and others do not~\cite{Gibbs}.
Since such
models cannot respect the Goldstone mechanism, their relevance to QCD is
questionable at best.

In this paper, we study the $N_f$-flavor four-fermi model with U(1)
chiral symmetry, sometimes called the Gross-Neveu model~\cite{GN}, in three
dimensions (ie. two space and one time) as
a function of chemical potential $\mu$. The action is
\begin{equation}
{\cal L}=\bar\psi_i(\partial{\!\!\! /}\,+\mu\gamma_0+m)\psi_i-
{g^2\over{2N_f}}[(\bar\psi_i\psi_i)^2-(\bar\psi_i\gamma_5\psi_i)^2],
\end{equation}
where $\psi, \bar\psi$ are four component spinors and a
sum on $i=1,\ldots,N_f$ is understood. For $\mu<\mu_c$
this model has a massless pion in the chiral
limit $m\to0$ \cite{RWP1} and can be studied by the leading order
$1/N_f$ expansion
\cite{RWP2}\cite{HKK1} as
well as by conventional computer
simulations.  Since the pathologies in QCD simulations are tied to $m_\pi$,
we can address the possibility that the presence of this massless mode is the
culprit in
past failures.  In addition, the model has composite mesons like QCD and has
sensible infra-red and ultra-violet properties.  The zero chemical
potential model
has a chiral transition as a function of its bare coupling which can be
analyzed using the $1/N_f$ expansion~\cite{HKK2}\cite{He}.
The second order phase transition coincides with a renormalization
group fixed point and defines an interacting continuum field theory~\cite{RWP2}
\cite{HKK2}.
It will prove interesting to simulate the U(1) four-fermi model to see
the physics of chiral symmetry restoration at work in a model with a
realistic, composite
pion.  The four-fermi model lacks two central features of QCD:  1.\ \ the
model does
not confine, and 2.\ \ its fermion determinant is real and non-negative
even when $\mu\neq 0$.

We shall see that the simulation study of the four-fermi model is completely
successful and physical, and in agreement with the predictions
of the $1/N_f$ expansion.  In
addition to the usual thermodynamic quantities such as the chiral condensate
$\langle{\overline\psi}\psi\rangle$, the fermion energy density $\epsilon$,
and the
fermion density $\rho$ itself, we will measure some spectroscopic features
of the model.  These will include the pion and fermion screening lengths, since
they are so closely tied to the physics issues of interest.
Recall how chemical potentials are
implemented and screening lengths are measured. Consider a symmetric $L^3$
lattice and label one of the axes $\tau$, ``temporal."
In the $+ \tau$ direction
assign a
factor exp($\mu$) to each such link in the action, and in the $- \tau$
direction
assign a factor exp($- \mu$).  These exponential factors favor quark
propagation in
the $+ \tau$ direction and when $\mu$ is small in units of the reciprocal
lattice spacing, ie. $\mu a\to0$, we have a useful,
well-behaved transcription of the chemical
potential to a discretized system~\cite{Kog}\cite{Has}.
Screening lengths are calculated in this environment by
calculating propagation with a source at $\tau_i = 0$ and a sink at $\tau_f =
\tau$.  The exponential fall-off with $\tau$ then gives an estimate of the
appropriate screening length, as will be illustrated through detailed
calculations in the text below.  The simplest expectation is that the pion
inverse screening length
should be zero in the chiral limit for small $\mu$ where chiral symmetry is
spontaneously broken, and as $\mu$ increases through a critical value
$\mu_c$, where
chiral symmetry is restored, the pion inverse screening length should increase
from zero non-analytically.
The fermion inverse screening length should decrease linearly
with $\mu$ and vanish at $\mu_c = m_f$, where $m_f$ is the value of the
fermion mass at $\mu=0$.  Interactions are expected to decrease $\mu_c$ below
this free field result, but at large $N_f$
the naive result should be accurate.  The result $\mu_c = m_f$ assumes that
the fundamental fermion is the lightest fermion in the model's mass
spectrum.  Although this result is expected
in three dimensions, it is not true in four-Fermi models in two dimensions,
where
there are ``kink" solutions and kink -- anti-kink bound states~\cite{DMR}.
Computer simulation
of the two dimensional model were successful in finding the kink solutions
and the subtle theoretical predictions were confirmed~\cite{KKW}.
In this study we shall confirm the naive expectation
$\mu_c = m_f$ with good control.  In fact, the $\mu$-dependence of the fermion
screening length will be consistent with,
\begin{equation}
m_f(\mu) = m_f (0) - \mu\ \ \ \ \ \ \ \ \ (\mu < \mu_c = m_f (0))
\end{equation}
which is also the free-field prediction.

The results of our $1/N_f$ calculation and lattice simulations
were very clear.  The
simulation gave the expected physical answers for both large $N_f$
($N_f=12$), and intermediate $N_f$ ($N_f=4$).
We worked near the chiral limit and found no
pathologies
when $\mu$ passed through the value $m_\pi /2$.  In fact, the algorithm was
well-behaved for all $\mu$ and computer simulations
in the immediate vicinity of
$\mu_c = m_f(0)$ were only limited by finite size effects of the expected
variety.
The chiral symmetry restoring transition was clear in all the observables
calculated
and excellent consistency was found between the various measurements.

Does this success help us understand the confused state of QCD simulations at
non-zero chemical potentials?  It certainly indicates that a massless pion
is {\sl not\/}
the culprit behind the failures encountered in QCD simulations.  The
compositeness of
the pion is also seen to be harmless.  As suggested by many authors in the
past \cite{Barb1}, the
complex nature of the QCD fermion determinant must be the source of the
problem.  The
quenched version of QCD ignores the complex part and possible rapid
variation in the
determinant and is probably pathological because of this, as illustrated in the
single-site U(1) model of Gibbs~\cite{Gibbs}.  A successful simulation of QCD
at
non-zero $\mu$
may require a wholly new algorithmic approach.  The present generation of
fermion
algorithms calculate the fermion propagator in a given gauge field
background,
and this intermediate step may not yield useful results
in QCD for $m_\pi/2 < \mu
< m_B/3$~\cite{KLS}.

This paper is organized as follows.  In Sec. II we define the
lattice version
of the U(1) four-fermi model and present some leading order large-$N_f$
results for the first order chiral symmetry restoring transition for
$\mu>0$
in the model, which
will be compared to computer simulations later in the paper.  In Sec. III we
consider
the computer simulations of the $N_f = 12$ and 4 models.
We shall find that the
usual bulk thermodynamic observables such as
$\langle{\overline\psi}\psi\rangle$,
energy densities and number densities successfully expose the chiral
transition.
In addition, spectroscopic measurements are equally clear and particularly
informative.  The pion inverse screening
length is essentially zero in the broken,
low $\mu$
phase, but is non-zero in the unbroken phase.  The fermion inverse
screening length
decreases linearly with $\mu$ in the broken phase and vanishes at the same
critical
point seen in the other observables, $\mu_c \simeq m_f(0)$,
the fermion mass observed at $\mu=0$.
The hybrid Monte Carlo algorithm we use is efficient for all $\mu$ and the
convergence of its
underlying conjugate-gradient routines to invert the lattice Dirac
operator does not deteriorate with increasing $\mu$.  Sec. IV includes a brief
summary and remarks on the lessons learned in this study.

\section{Lattice Formulation of the Gross-Neveu Model}

The lattice action for the bosonized Gross-Neveu model with U(1) chiral
symmetry is
\begin{eqnarray}
S=\sum_{i=1}^{N_f/4}\biggl[\sum_{x,y}\bar\chi_i(x){\cal M}_{x,y}\chi_i(y)
&+&{1\over8}\sum_x\bar\chi_i(x)\chi_i(x)
\Bigl(\sum_{<\tilde x,x>}\sigma(\tilde x)+i\varepsilon(x)
\sum_{<\tilde x,x>}
\pi(\tilde x)\Bigr)\biggr] \nonumber \\
&+&{N_f\over8g^2}\sum_{\tilde x}
(\sigma^2(\tilde x)+\pi^2(\tilde x)).
\end{eqnarray}
Here, $\chi_i$ and $\bar\chi_i$ are complex Grassmann-valued staggered
fermion fields defined on the lattice sites, the auxiliary scalar and
pseudoscalar fields $\sigma$ and $\pi$ are defined on the dual lattice
sites, and the symbol $<\tilde x,x>$ denotes the set of 8 dual sites
$\tilde x$ adjacent to the direct lattice site $x$.
The auxiliary fields only appear to quadratic order, and can be
integrated out to recover a lattice action in terms of fermion
fields only (bosonization is also possible for the continuum Lagrangian
(1.1)). The parameter $N_f$
will turn out to be the number of physical fermion species described by
the numerical simulation, and must be an integer multiple of 4. The
fermion kinetic operator ${\cal M}$ is given by
\begin{equation}
{\cal M}_{x,y}={1\over2}\Bigl[\delta_{y,x+\hat0}e^\mu-
\delta_{y,x-\hat0}e^{-\mu}\Bigr]+{1\over2}\sum_{\nu=1,2}\eta_\nu(x)
\Bigl[\delta_{y,x+\hat\nu}-\delta_{y,x-\hat\nu}\Bigr]+m\delta_{y,x},
\end{equation}
where $m$ is the bare fermion mass, $\mu$ is the chemical potential, and
$\eta_\nu(x)$ are the Kawamoto-Smit phases
$(-1)^{x_0+\cdots+x_{\nu-1}}$. The symbol $\varepsilon(x)$ denotes the
alternating phase $(-1)^{x_0+x_1+x_2}$.

As described in refs.~\cite{HKK1}\cite{HKK2},
the model (2.1) can be recast in terms
of fields $q(Y)$ defined on a lattice of twice the spacing, where the
field $q$ has explicit spinor and flavor indices. In momentum space the
kinetic part of the action reads
\begin{eqnarray}
S_{kin} &=& \sum_i\int {d^3k\over(2\pi)^3}\sum_{\nu=1,2}{i\over2}\left[
            \bar q_i(k)(\gamma_\nu\otimes1\kern-4.5pt1_2)q_i(k)\sin2k_\nu+
            \bar q_i(k)(\gamma_4\otimes\tau_\nu^*)q_i(k)(1-\cos2k_\nu)
            \right]\nonumber\\
        & & +{1\over2}\biggl[
          \bar q_i(k)(\gamma_0\otimes1\kern-4.5pt1_2)q_i(k)[i\sin2k_0\cosh\mu
            +(1+\cos2k_0)\sinh\mu]\\
        & & +\bar q_i(k)(\gamma_4\otimes\tau_3^*)q_i(k)[i(1-\cos2k_0)\cosh\mu
            +\sin2k_0\sinh\mu]\biggr]
            +m\bar q_i(k)(1\kern-4.5pt1_4\otimes1\kern-4.5pt1_2)q_i(k),
            \nonumber
\end{eqnarray}
with $\tau_i$ the Pauli matrices, which act on the 2 component flavor
degrees of freedom, and
\begin{equation}
\gamma_\nu=\left(\matrix{\tau_\nu&\cr&-\tau_\nu\cr}\right);\;\;\;
\gamma_0  =\left(\matrix{\tau_3&   \cr&-\tau_3 \cr}\right);\;\;\;
\gamma_4  =\left(\matrix{&-i1\kern-4.5pt1_2\cr
          i1\kern-4.5pt1_2&\cr}\right);\;\;\;
\gamma_5  =\left(\matrix{&1\kern-4.5pt1_2\cr 1\kern-4.5pt1_2&\cr}\right).
\end{equation}
The momentum integral extends over the range $k\in(-\pi/2,\pi/2]$. On
a finite system the integral is replaced by a sum over $L/2$ modes,
where $L$ is the number of lattice spacings in a given direction. In
the large wavelength limit $k\to0$ we recover the standard Euclidean
continuum form $\bar q_j(\partial{\!\!\! /}\,+\mu\gamma_0+m)q_j$, where
$j$ now runs from 1 to $N_f/2$.

The interaction part of the action can be rewritten as
\begin{equation}
S_{int}=\sum_Y\hat\sigma(\tilde Y)\bar q_i(Y)
       (1\kern-4.5pt1_4\otimes1\kern-4.5pt1_2)q_i(Y)
               +\hat\pi(\tilde Y)\bar q_i(Y)
                (i\gamma_5\otimes1\kern-4.5pt1_2)q_i(Y)
           +O(a),
\end{equation}
where $\hat\sigma(\tilde Y)$, $\hat\pi(\tilde Y)$ denote the sum of
the eight scalar fields associated with the site $Y$ on the blocked
lattice (see \cite{HKK2} for full details), and $a$ is the lattice spacing,
which has been set to unity in Eqns. (2.1-3). The $O(a)$ terms are
non-covariant and flavor non-singlet -- if we used a formulation in
which the scalar fields were defined on the direct lattice sites, then
such non-covariant terms would appear at $O(a^0)$~\cite{CER}.

The lattice action (2.1) has a global vector U($N_f/4$) symmetry
\begin{equation}
\chi_i\mapsto\Omega_{ij}\chi_j\;\;;
\;\;\bar\chi_i\mapsto\bar\chi_j\Omega^\dagger_{ji}\;\;;
\;\;\;\;\;\;\;\;\;\Omega\in{\rm U}(N_f/4),
\end{equation}
and, in the chiral limit $m\to0$, a global ``axial'' U(1) symmetry
\begin{equation}
\chi\mapsto e^{i\alpha\varepsilon(x)}\chi\;\;;
\;\;\bar\chi\mapsto\bar\chi e^{i\alpha\varepsilon(x)}\;\;;
\;\;\phi\equiv(\sigma+i\pi)\mapsto e^{-2i\alpha}\phi.
\end{equation}
In the $q$-basis the rotation (2.7) reads:
\begin{equation}
q\mapsto\exp i\alpha(\gamma_5\otimes1\kern-4.5pt1_2)q\;\;;\;\;
\bar q\mapsto\bar q\exp i\alpha(\gamma_5\otimes1\kern-4.5pt1_2).
\end{equation}
The U(1) symmetry is spontaneously broken by the condensate
$\langle\bar\chi\chi\rangle$ (or equivalently $\langle\sigma\rangle$):
for $m\not=0$ this
direction of symmetry breaking is picked out.

The action (2.1) may be simulated using a hybrid Monte Carlo algorithm
{}~\cite{DKPR},
in which the Grassmann fields are replaced by complex commuting
pseudo-fermion fields $\psi(x)$ governed by the action
\begin{equation}
S=\sum_{xy}\sum_{ij=1}^{N_f/4}{1\over2}\psi_i^*(x)(M^\dagger
M)^{-1}_{xyij}\psi_j(y)+{N_f\over8g^2}\sum_{\tilde x}\biggl(\sigma^2(\tilde
x)+\pi^2(\tilde x)\biggr),
\end{equation}
where
\begin{equation}
M_{xyij}={\cal M}_{xy}\delta_{ij}+\delta_{xy}\delta_{ij}
{1\over8}\sum_{<\tilde x,x>}[\sigma(\tilde
x)+i\varepsilon(x)\pi(\tilde x)].
\end{equation}
Integration over the $\psi$ fields yields the functional measure
$\det(M^\dagger M)$. Note that the kinetic part of $M$, ${\cal M}$, is
strictly real even for $\mu\not=0$, and that the complex part
$\sigma+i\varepsilon\pi$ is diagonal. Thus, schematically,
\begin{eqnarray}
\det(M^\dagger M)&=&\det M^*\det M\nonumber\\
                 &=&\det({\cal M}+\sigma+i\varepsilon\pi)
                   \det({\cal M}+\sigma-i\varepsilon\pi).
\end{eqnarray}
Since each matrix ${\cal M}$ effectively describes $N_f/2$ fermion
species, we conclude that the hybrid Monte Carlo simulation describes
a system of $N_f$ fermion species, $N_f/2$ having positive chiral
charge and $N_f/2$ negative. To be more precise, the full symmetry of
the lattice model in the continuum limit is
${\rm U}(N_f/2)_V\otimes{\rm U}(N_f/2)_V\otimes{\rm U}(1)_A$ rather
than the ${\rm U}(N_f)_V\otimes{\rm U}(1)_A$ we would obtain if all
species transformed identically with respect to chiral rotations. At
non-zero lattice spacing, of course, the symmetry group is smaller as
discussed above:
${\rm U}(N_f/4)_V\otimes{\rm U}(N_f/4)_V\otimes{\rm U}(1)_A$.
In all cases, of course, it is the ${\rm U}(1)_A$ symmetry which is
broken, either spontaneously by the dynamics of the system, or
explicitly by a bare fermion mass.

It was found that the performance of the hybrid Monte Carlo procedure
could be optimized in two ways: firstly, as described in
\cite{HKK1}\cite{HKK2},
by tuning the effective number
of fermion species $N_f^\prime$ used in the guidance part of the
program (ie. during the integration of the equations of motion
along a microcanonical trajectory) to maximize the
acceptance rate of the Monte Carlo procedure for fixed timestep
$\Delta\tau$; and secondly by choosing
the trajectory length
$\tau$ at random from a Poisson distribution of mean
$\bar\tau$. This second refinement, which guarantees ergodicity, was found to
dramatically decrease
autocorrelation times~\cite{Nf2QED}.

In our specific application we found that the choice $\Delta\tau = 0.01$ was
adequate -- the acceptance rate in the Monte Carlo procedure was always high
(better than 80\%) while the algorithm sampled configuration space with good
speed and efficiency.  For the $N_f = 4$ model the acceptance rate was improved
by taking $N^\prime_f$ slightly larger, typically 4.05, although the
``best" choice of
$N^\prime_f/N_f$ certainly depends on lattice sizes ($16^3$ in our case)
and couplings
($0 \mathrel{\mathpalette \vereq <} 1/g^2
\mathrel{\mathpalette \vereq <} 1.0)$
and bare fermion masses (0.01 in lattice units).  The trajectory lengths were
chosen from a Poisson distribution with $\overline{\tau}$ typically between 1
and 2.  In the immediate vicinity of the critical $\mu$, a larger
$\overline{\tau}$ probably would have been better, but we had no trouble
obtaining good data with modest error bars from runs of several hundred
trajectories.  The data and error bars will be presented in tables in the next
section.  We used identical parameters when the $N_f = 12$ model was
simulated except $N_f^\prime$ was now tuned typically to 12.12,
for better acceptance rates.

As well as measuring the expectation value of the scalar field
$\langle\sigma\rangle$ in the simulation, which for our purposes serves as an
order parameter, we also monitored the chiral condensate
$\langle\bar\chi\chi\rangle$, the energy density $\langle\epsilon\rangle$,
and the fermion
number density $\langle\rho\rangle$, which are defined as follows:
\begin{eqnarray}
-\langle\bar\chi\chi\rangle&=&{1\over V}{\rm tr}S_F={1\over V}
\langle{\rm tr}M^{-1}\rangle,\nonumber\\
\langle\epsilon\rangle=-{1\over V_s}{{\partial\ln Z}\over{\partial\beta}}
&=&{1\over V}{\rm tr}\partial_0\gamma_0S_F=
{1\over2V}\langle\sum_x
e^\mu M^{-1}_{x,x+\hat0}-e^{-\mu}M^{-1}_{x,x-\hat0}\rangle,\\
\langle\rho\rangle=-{1\over{V_s\beta}}{{\partial\ln Z}\over{\partial\mu}}
&=&{1\over V}{\rm tr}\gamma_0S_F=
{1\over2V}\langle\sum_x
e^\mu M^{-1}_{x,x+\hat0}+e^{-\mu}M^{-1}_{x,x-\hat0}\rangle.\nonumber
\end{eqnarray}
Here $V_s$ is the spatial (ie. two-dimensional) volume, $\beta$ the
inverse temperature, and $V=V_s\beta$ the overall volume of spacetime.
The final expression in each case is the quantity measured in the
simulation, using a noisy estmator to calculate the matrix inverses.

Finally, as well as results from numerical simulations, we can examine
the action (2.1) using the $1/N_f$ expansion. To leading order, this
corresponds to the saddle point solution of the path integral, or
equivalently to mean field theory. We can solve for the expectation
value of the scalar field $\langle\sigma\rangle$ using the gap
equation~\cite{RWP2}\cite{HKK1}:
\begin{equation}
\langle\sigma\rangle=-g^2\langle\bar\chi\chi\rangle={g^2\over V}{\rm tr}S_F.
\end{equation}
To leading order the gap equation consists of a single tadpole.
Using the form for the kinetic term (2.3), we solve self-consistently to
get an equation for $g^2$ in terms of the dynamical fermion mass
$m_f=\langle\sigma\rangle+m$. In the thermodynamic limit $L\to\infty$,
\begin{equation}
{1\over g^2}={8\over{2\pi}^3}{m_f\over{m_f-m}}
\int_{-\pi/2}^{\pi/2}
{{d^3k}\over{
{1\over2}\left\{1-\cos2k_0\cosh2\mu-i\sin2k_0\sinh2\mu\right\}
+\sum_{\nu=1,2}\sin^2k_\nu+m_f^2}},
\end{equation}
Note that the $O(a)$ terms in the interaction (2.5) make no contribution
at this order. Eqn. (2.14) can
be reduced to a one-dimensional integral
and then evaluated numerically in the limit $\mu\to0$~\cite{HKK2}: however for
$\mu\not=0$ it is more efficient to simply evaluate the sum over
lattice momenta on finite systems explicitly.
Note that antiperiodic
boundary conditions must be specified on the fermion fields in the
timelike direction; we chose to apply periodic boundary conditions in
the two spacelike ones.

In Figs. 1 - 5 we show the large-$N_f$ predictions
from Eqn. (2.14) evaluated on  a $16^3$ lattice. In Fig. 1 we show
$\langle\sigma\rangle$ vs. $1/g^2$. We have chosen bare fermion
masses of $m=0.05, 0.01\ {\rm and}\ 0.00$.  This range of masses shows the
sensitivity of the finite lattice results to $m$, gives the theoretically
interesting chiral limit $m=0.00$ and includes $m=0.01$
to compare quantitatively
to the simulation results obtained
from the hybrid Monto Carlo algorithm at $N_f
= 12\ {\rm and}\ 4, m = 0.01$.
A
chiral symmetry restoring symmetry point at $1/g^2_c$ slightly larger than 1.00
is found consistent with the critical index $\beta_{mag} = 1.0$~\cite{HKK2}.
The curves at $m=
0.01$ and
0.05 smooth the transition out in the expected fashion. Only the $m=0.00$ curve
shows evidence of finite volume rounding (this was checked by evaluating the
gap equation on larger lattices).
Note that
$\langle\sigma\rangle$ is also the fermion dynamical mass $m_f$
in the chiral limit, so $m_f$ is an equally good order parameter for the
transition. Since $m_f$ is an inverse correlation length, the critical
exponent
$\nu = 1.0$ follows from the figure.
In Fig. 2 we show $\langle\sigma\rangle$
vs. $\mu$ for a bare coupling $1/g^2=0.5$ deep within
the broken symmetry phase.  This coupling
will be simulated in the next section.  We note that $\langle\sigma\rangle$ is
essentially unaffected by $\mu$ until the immediate vicinity of the transition
where $\langle\sigma\rangle$ jumps to zero through a first order phase
transition.  (Recall that in mean field theory a
model can have a phase transition in a
finite volume.)  The rounding of the curves near $\mu=\mu_c$ is a finite volume
effect.
The critical $\mu_c$ is approximately 0.73 which compares
rather
well with the dynamical fermion mass in the chiral limit at $1/g^2= 0.5$
and $\mu = 0.0, m_f =
0.84$,
seen in Fig. 1.  The discrepancy between $\mu_c$ and $m_f$ is simply due to the
discreteness of the lattice sums approximating continuum integrals.
In the continuum limit $g\to g_c$ (ie. $m_f\to0$), the discrepancy would
approach zero. It is interesting to note that if the plot of $m_f$ vs.
$1/g^2$ in Fig. 1 is extrapolated back from $1/g_c^2$ using the
assumption of linearity (ie. $\beta_{mag}=1$), then the resulting $m_f$
at $1/g^2=0.5$ is in much closer agreement.
The $m=0.01$ and 0.05
curves in Fig. 2 demonstrate that these bare masses do not distort the first
order symmetry restoring transition significantly.  Since the mass ratios
$m/m_f$
are 0.012 and 0.058, approximately,
in the two cases and since the transition is
strongly first order, this result is expected.
In Fig. 3 we show the lattice gap equation prediction for $\sigma$ vs.
$\mu$ at $m=0.01$ on a $16^3$ lattice for values of $1/g^2$ ranging from 0.5 to
0.8, showing
the effect of finite volume rounding becoming more pronounced as the
physical lattice spacing is reduced. By $1/g^2=0.8$ the transition has
become much less sharp.
In Figs. 4 and 5 we show the
fermion number density $\rho$ and the fermion energy density $\epsilon$ plotted
against $\mu$ in the chiral limit for three choices of couplings in the broken
phase, $1/g^2= 0.5, 0.7\ {\rm and}\ 0.9.$  Also shown in Fig. 4 is the
expected continuum result for fermions in the unbroken phase,
$\rho=\mu^2/\pi$ (we have rescaled the numerical results for $\rho$ and
$\epsilon$ by a factor of two over the definitions (2.12))~\cite{HKK1}.
For $1/g^2=0.9$ the transition is hard to identify.
As we approach the bulk critical
point $1/g^2_c \approx 1.0$ on the finite system the signals of the phase
transition become less and less dramatic.  Larger lattices are necessary to
obtain a quantitative picture of the transition when $1/g^2$ is chosen to be
0.9, close to the bulk continuum limit $1/g^2_c\approx 1.0.$

\section{Simulation Results}

We simulated both the $N_f = 12$ and $N_f = 4$ models at various couplings and
chemical potentials. In all cases a $16^3$ lattice was used, with a bare
fermion mass $m=0.01$.
As explained above, due to fermion ``doubling," the
$N_f = 12(4)$ model
corresponds to $N_{f{\rm latt}}=3(1)$
lattice species.  We simulated the $N_{f{\rm latt}}=3$ model
because it
should compare well with the results of the $1/N_f$ expansion
(if its underlying assumptions hold in the model),
and observables should be large with relatively
modest fluctuations.  The $N_{f{\rm latt}}=1$
model is interesting because it could show
qualitative deviations from the large-$N_f$ results;
its observables will fluctuate
more intensely and it will present a numerical challenge closer to that
of two- or four-flavor
QCD.  The observables and measurement techniques were discussed above.
They have also been used in our past studies of four-fermi models, so we refer
the reader to those references rather than repeat standard material~\cite{HKK1}
\cite{HKK2}.
The ``new" measurements we have done concern the model's spectroscopy.  These
observables are particularly revealing and we will discuss them at greater
length.

Using point sources, we calculated the fermion propagator,
$G_+(\vec{x},t)$, with chemical potential $\mu$ and,
$G_-(\vec{x},t)$, with chemical potential $-\mu$.  Then we
formed a zero momentum fermion propagator
\begin{equation}
P_f(t) = \sum_{\vec{x}} G_+(\vec{x},t),
\end{equation}
and anti-fermion propagator
\begin{equation}
P_{\bar{f}}(t) = \sum_{\vec{x}}
G_-(\vec{x},t).
\end{equation}
Of course, $G_+(\vec{x},t) = G_-(\vec{x},t)$
for the zero chemical potential case.  The composite
pion and sigma propagators
are,
\begin{equation}
P_\pi(t) =
\sum_{\vec{x}}G_+(\vec{x},t)G_-^\dagger
(\vec{x},t),
\end{equation}
\begin{equation}
P_\sigma(t) =
\sum_{\vec{x}}(-1)^{x+y}G_+
(\vec{x},t)G_-^\dagger(\vec{x},t),
\end{equation}
in analogy with the treatment using staggered fermions in four
dimensions~\cite{SB}.
We also calculated propagators for the
auxiliary fields
in Eq. (2.1), $\pi$ and $\sigma$,
\begin{equation}
P_\pi (t) = \sum_{\vec{x},t^\prime}
\pi(\vec{x},t^\prime )\pi
(\vec{x},t^\prime + t),
\end{equation}
and
\begin{equation}
P_\sigma (t) =
\sum_{\vec{x},t^\prime}
[\sigma(\vec{x},t^\prime)
\sigma(\vec{x},t^\prime
+ t) - \overline{\sigma}^2].
\end{equation}
Here, $\overline{\sigma}^2$ is the square of the average of the $\sigma$ field.

After calculating averages of the above propagators and their covariance
matrices
(see \cite{SB} for fitting techniques), we fit the various propagators to the
following
functional forms. For zero chemical potential:
\begin{equation}
P_f(t) = A[e^{-m_f t} - (-1)^t e^{-m_f(T-t)}],
\end{equation}
where $T$ is the temporal extent of the lattice, for the fermion and
\begin{equation}
P_\pi(t) = A [e^{-m_\pi t} + e^{-m_\pi (T-t)}],
\end{equation}
for the auxiliary field, $\pi$.  Similar functional forms are chosen for the
anti-fermion and $\sigma$ respectively.
For the non-zero chemical potential case, we chose the forms,
\begin{equation}
P_f(t) = A[e^{-m^f_ft} - (-1)^t e^{-m^b_f (T-t)}]
\end{equation}
\begin{equation}
P_\pi (t) = A[e^{-m^f_\pi t} + e^{-m^b_\pi (T-t)}].
\end{equation}
Since the meson has zero fermion number, we expect that $m^f_\pi = m^b_\pi$.

Those parameters which minimize correlated $\chi^2$ were chosen as the best
fitted
values.  We used the CERN mathematical library routine, MINUIT, as a
minimization
program.
The error bars quoted refer to the necessary parameter changes for
a change of $\chi^2$ by one.
Only two of our spectrum calculations yielded useful, accurate mass
estimates.  They were
the fermion and the auxiliary field pion masses. Luckily, these are the
two quantities
most closely related to chiral symmetry and its restoration
at non-zero chemical
potential.  It would have been interesting to calculate the $\sigma$ mass,
but the
fluctuations in our limited data sets made that impossible.

Now let's turn to the data.  The $N_{f{\rm latt}}=3$
data for $m_f, m_\pi$, the vacuum expectation value
$\langle\sigma\rangle$, which we shall denote $\sigma$ for simplicity
in this section,
and the action $S$ are given in Table I for $\mu = 0$ and
couplings
$1/g^2$ ranging from 0.50 to 1.0.  The order parameter $\sigma$ and $m_f$
are almost
identical as they should be at large $N_f$.
We show a plot of $\sigma$ vs.
$1/g^2$ in Fig. 6.  Clearly chiral symmetry is broken for $1/g^2
\mathrel{\mathpalette \vereq <} 0.90-1.00$. Also shown is the prediction
of the lattice gap equation from Fig. 1.
The agreement is $O$(10\%) or better, and
is very satisfactory.  It is our first indication that
the $1/N_f$ expansion is
practical and obtains the correct physics of these models.  We note from
the table
that the mass of the pion is ``small" and quite insensitive to $1/g^2$.
Since chiral
symmetry is broken over this range of couplings, it should be that the
pion's mass is
nonzero only because of the explicit symmetry breaking provided by the
small bare
fermion mass, $m = 0.01$.  We will obtain good evidence for this
interpretation of the
data when we consider the model at nonzero chemical potential.

We studied the model next at $1/g^2=0.50$ as a function of $\mu$ to see how a
nonzero chemical potential restores chiral symmetry at a critical point.
$1/g^2=
0.50$ is a good place to look first because it is deep in the broken
symmetry phase,
but not too far from the bulk critical point $1/g^2\approx 1.00.$  As long
as $1/g^2
\approx 0.50$ is in the scaling window of the critical point we can extract
continuum
physics from these simulations by standard methods.  Our goals here are more
modest --
we wish to turn up $\mu$ and see if the conventional picture of symmetry
restoration emerges.  The simulation data for $1/g^2=0.50$ and $\mu$
ranging from
0.50 to 0.85 is given in Table II.  The table includes data for $m_f$ (the
fermion
mass), $m_\pi$ (the pion mass), $\rho$ (the fermion number density), $\sigma$
(the vacuum
expectation value of the $\sigma$ field), $S$ (the action), and $\epsilon$
(the energy
density).  Since the fermion mass at $\mu = 0.0$ is $m_f =0.746(2)$
according to
Table I, we expect naively a chiral symmetry restoring transition at $\mu_c =
0.746(2)$.  This result is beautifully reproduced by the simulation.  In
Fig. 7 we
plot the order parameter $\sigma$ vs. $\mu$ for fixed $1/g^2=0.50$, and we see
restoration of the symmetry at $\mu_c =0.725(25)$.  The curve is very
abrupt and we
suspect, naturally, that a simulation on a larger lattice at smaller bare
fermion
mass $m$ would show a first order discontinuous transition.  The curve is
also quantitatively very similar to the predictions of the lattice gap
equation which is also shown -- once again, the discrepancy is $O(10\%)$ and
may presumably be ascribed to $O(1/N_f)$ corrections.

As discussed in the introduction,
one of the
purposes of
studying this model was to verify that the lattice formulation, given a proven
algorithm, obtains the correct physics at non-zero chemical potential even in
a model
with a Goldstone pion.  Quenched QCD simulations have pathologies when
$\mu$ approaches
$m_\pi /2$, and although the chiral restoring transition
is expected at $\mu_c =
m_B/3,$ one-third the mass of the nucleon, there is little numerical
evidence for this~\cite{Barb2}\cite{DK}\cite{KLS}.
Many reasons have been proposed in hindsight for this catastrophe and
several of them
hinge on the presence of a Goldstone pion in the theory's spectrum.  In the
simulations here $m_\pi  /2 \approx0.09$ while the expected transition is
at $\mu_c =
m_f =0.74(2)$.  In QCD simulations the two mass scales, $m_\pi  /2$ and
$m_B /3$,
are actually quite close in numerical simulations performed to date, and this
has further
clouded the situation.  In our U(1) four-fermi model, we can
simulate
the model very close to the chiral limit in the sense that the explicit
breaking is much smaller than the dynamical breaking (ie. bare mass
$m\ll m_f$, the dynamical mass).
Fig. 7 and the
rest of the data in Table II show that $m_\pi  /2$
is not a point of any special
significance and the simulation algorithm gives the expected $\mu_c =
0.74(2)$ nicely.
Additional simulations show that the value $m=0.01$ is not crucial to these
conclusions, and the chiral limit $m \rightarrow 0$ does not contain
surprises even
when $\mu \neq 0$.  For example, in Fig. 8 we show $\sigma$ vs. $m$ for $1/g^2
=0.50$
and $\mu =0.50$.  Unlike the tortuous situation in quenched QCD where plots of
$\langle{\overline\psi}\psi\rangle$ vs. $m$ have strong downward curvature for
$\mu\mathrel{\mathpalette \vereq >} m_\pi /2$ \cite{Barb1},
the extrapolation to the chiral limit
here is essentially linear and without surprises.
Simulations at larger values of $m$ would have been as clear as the $m = 0.01$
case studied here in detail.

The transition at $\mu_c =0.74(2)$ is seen equally well in the fermion
number and
energy densities (Fig. 9), and the action itself (Fig. 10).  These
curves
compare well with the analytic large-$N_f$ results presented in Sec. II above.
The spectroscopy of the model and its dependence on $\mu$ is particularly
interesting.  A recent study of quenched QCD \cite{KLS}
showed that in the limited range $0 < \mu < m_\pi /2$,
the baryon mass as defined through the exponential falloff of a Euclidean
propagator
as in Eq. (3.7) decreased linearly with $\mu$ and would vanish by linear
extrapolation at $\mu_c = m_B/3$ as expected.  Unfortunately, the quenched
simulation algorithm suffers from slow convergence and large fluctuations
for $\mu\mathrel{\mathpalette \vereq >}m_\pi
/2$, so
nothing is known quantitatively in the ``forbidden" region, $m_\pi /2 < \mu
< m_B
/3.$  In addition, the quenched QCD simulation showed that the pion mass,
as defined
through the exponential fall-off of a propagator as in Eq. (3.8), is
insensitive to
$\mu$ for $\mu\mathrel{\mathpalette\vereq <}m_\pi /2$.
This is another sensible
result which could not be
confirmed at larger $\mu$ due to the pathologies of the quenched simulation.
In our four-fermi model the analogous
calculations are successful for all $\mu$.  In Fig. 11 we show the fermion
mass as
obtained from Eq. (3.7).  Up to modest and expected finite size effects
which reduce
the fermion mass estimates in the vicinity of $\mu_c$, the calculation is
successful,
and gives a critical chemical potential near 0.74(2), although larger
lattice studies
and a systematic analysis of finite size effects would be necessary to obtain
a quantitative prediction.  Also in Fig. 11 we show the pion mass
$m_\pi$,
calculated using
Eq. (3.8), as a function of $\mu$.  The pion mass exhibits no $\mu$
dependence until
we reach the vicinity of $\mu_c =0.74(2)$ where it jumps up indicating chiral
symmetry restoration.  We wanted to measure the $\sigma$ mass over the same
range,
but its propagator was noisier than the pions and quantitative estimates
were not
achieved.
We had hoped to verify that the pion and the sigma are degenerate and
heavy for $\mu >0.74(2)$, indicating chiral symmetry restoration.  Although
from this
perspective our calculations were only partially successful, they gave decisive
physical answers expected of a Goldstone particle as we pass through a chiral
symmetry restoring transition.

Next we repeated these measurements at $1/g^2=0.80$ in the $N_{f{\rm
latt}}=3$ model in
order to be
closer to the model's continuum limit.
At $1/g^2=0.80$ the $\mu = 0$ value of the fermion mass
is $m_f =
0.295(3)$.  On the basis of the analytic large-$N_f$ results shown in
Figs. 3,4
and 5, we
anticipate that the transition will be harder to identify since finite size
effects
are more severe and smooth the transition considerably on a $16^3$ lattice.
We show
the data for $1/g^2=0.80$ in Table III.  It is organized just as Table II
was.  In Figs.
12 - 14 we show $\sigma, \rho, \epsilon$ and $S$ plotted vs. $\mu$.  These
figures show
the same features as the large-$N_f$ calculation,
and indicate that the
transition is in the vicinity of $\mu_c =0.295(3)$.  There is no evidence
whatsoever
for pathological behavior at $\mu=m_\pi /2 \approx0.11(1)$.  The transition is
shown with
greater clarity in the model's spectroscopy.  In Fig. 15 we see that the
dynamical fermion mass decreases, up to the expected finite size effects,
linearly
with $\mu$ and vanishes when $\mu$ becomes 0.25(5).
Similarly, the pion mass is insensitive to $\mu$ until $\mu=0.30(2)$,
where it increases noticeably.  We learn that the
model's spectroscopy is a more sensitive guide to the chiral restoration
transition
than traditional bulk thermodynamic quantities.  In light of our recent
work on
quenched QCD~\cite{KLS}, this result comes as
no surprise, but it reinforces the
strategy we are
taking in the four dimensional gauge model.

Next we turn to the $N_{f{\rm latt}}=1$
model to simulate a case where fluctuations are
expected to
be more significant, the number of flavors is more realistic and the
large-$N_f$
expansion may not be as good a guide.  The $\mu = 0$ data is collected in
Table IV.
We notice, as plotted in Fig. 16, that the order parameter $\sigma$ and
dynamical
fermion mass $m_f$ disagree slightly deep in the broken symmetry phase, but are
otherwise in good agreement.  Since these two quantities are identical at
large-$N_f$, we
see signs of $1/N_f$ corrections here, but they are not numerically significant
near the
transition.  We will investigate the theory at non-zero $\mu$ at both $1/g^2
=0.60$ and
0.70.  The $1/g^2=0.60$ simulation
is quite far in the broken phase and should be
relatively
decisive while the $1/g^2=0.70$ simulation will be more strongly affected by
fluctuations.  In Figs. 17 - 19 we show $\sigma , \rho ,\epsilon$ and $S$
plotted against
$\mu$ using the data of Table V.
In this case the plots show only qualitative agreement with the predictions
of the $1/N_f$ expansion at leading order; $O(1/N_f)$ corrections are
numerically much more significant.
We expect a transition near the value of the
dynamical fermion mass at $1/g^2= 0.60$, ie, $m_f =0.475(5)$,
and the figures are in fine
agreement with that.  In addition the fermion and pion masses, Fig. 20,
show
the transition almost as clearly and quantitatively as they did in
the $N_{f{\rm latt}}=3$,
$1/g^2=0.50$ case.  Finally in Table VI and Figs. 21-24 we show the analogous
quantities for the
$N_{f{\rm latt}}=1$ theory at $1/g^2=0.70$.  Although the bulk thermodynamic
quantities experience considerable rounding, the results are consistent
with a critical chemical potential $\mu_c=0.32(1)$, as predicted by the
value of
the dynamical fermion mass at $1/g^2=0.70$.
Once again, the spectroscopic quantities in Fig. 24 provide more quantitative
information.
All in all, the simulations are successful in each case and do not
suffer
from the pathologies affecting quenched QCD.

\section{Concluding Remarks}

The success and clarity of these simulations shows that the Hybrid Monte Carlo
algorithm (and hence presumably the closely related Hybrid Molecular
Dynamics algorithm, suitable for values of $N_f$ which are not a multiple of
4)
are completely reliable for this class of
fermion field theories in which there is a massless pion in the chiral limit.
A chiral symmetry restoring phase transition, probably first order,
is found for a critical value of
the chemical potential $\mu$.
Screening length calculations proved to be particularly illuminating and the
expected physics of chiral symmetry restoration at the critical chemical
potential emerged. The spectroscopic data may well prove to be the most
accurate means of determining the critical
chemical potential on finite systems.
The predictions of the large $N_f$ expansion proved to be a good guide into
the physics of these four fermi models even when $N_f$ assumed modest values.
Systematic effects are of the expected form, and we have no reason to suspect
they could not be brought under complete control given sufficient computer
time.

One of our primary motivations for this work was to narrow down the
source of difficulties in simulations of lattice QCD. We certainly have
demonstrated that the presence of a massless pion in the theory's spectrum
is {\it not} the source of those difficulties.
By default it must be the complex
nature of the QCD action at nonzero $\mu$ that is the culprit. The quenched
version of QCD ignores the fermion determinant and this
omission apparently amounts
to a qualitative error when the chemical potential is larger than
half the pion mass \cite{KLS}. In four
fermi models the fermion determinant is real and non-negative and the Euclidean
theory has a probabilistic interpretation. In addition, the fermion
propagator is well-behaved for all $\mu$. These ingredients then allow the
Hybrid algorithms to be successful in four fermi models despite being
inapplicable
for QCD. It may well be that Langevin algorithms \cite{P}
can simulate QCD directly at nonzero
$\mu$, but that is far from clear at this time (Langevin algorithms
are known to correctly evaluate certain complex integrals, and certain
physical systems with complex actions, but their successes and failures
are difficult to anticipate in advance). It will probably require
greater insight into the numerics of the complex fermion determinant
before a truly trustworthy, first principles simulation of QCD in an
environment rich in baryons is possible.

\section{Acknowledgements}
JBK is partially supported by the National Science Foundation under
the grant NSF PHY92-00148. SK is supported by the
U.S. Department of Energy, Contract No. W-30-109-ENG-38.
SJH was supported in part by a CERN fellowship, and in part by a
PPARC Advanced fellowship. The computer simulations were done on the Cray C90's
at NERSC and PSC. We thank these computer centers for their help.

\begin{table}
\caption{\
$N_{f{\rm latt}}=3$
data on $16^3$ lattice at $\mu = 0.0$.  The columns give the
coupling
$1/g^2$, the fermion mass $m_f$, the pion mass $m_\pi$, the vacuum
expectation value
of the $\sigma$ field $\sigma$, the action $S$.}
\label{1.}
\begin{tabular}{dddcc}
$1/g^2$    & $m_f$ & $m_\pi$ & $\sigma$ & $S$\\
\tableline
1.00 & .102(2) & .22(1) & .094(2) & .363(2)\\
.95 & .133(2) & .22(3) & .128(2) & .390(2)\\
.90 & .182(2) & .19(3) & .176(3) & .428(2)\\
.85 & .233(3) & .19(3) & .231(3) & .475(3)\\
.80 & .295(3) & .21(3) & .294(3) & .535(3)\\
.75 & .359(2) & .18(1) & .365(3) & .613(3)\\
.70 & .430(2) & .18(2) & .438(4) & .707(3)\\
.65 & .503(3) & .18(2) & .519(4) & .824(4)\\
.60	& .573(3) & .18(1)	& .604(4) & .969(5)\\
.55 & \underline{\ \ \ \ \ } & \underline{\ \ \ \ \ } & .695(4) & 1.14(1)\\
.50 & .746(2) & .18(1) & .795(3) & 1.36(1)
\end{tabular}
\end{table}

\begin{table}
\caption{\
$N_{f{\rm latt}}=3$
data on $16^3$ lattice at coupling $1/g^2=0.50$ for various
chemical
potentials $\mu$.  Same notation as Table I but $\rho$ is the fermion
density and
$\epsilon$ is the energy density.}
\label{2.}
\begin{tabular}{dcccccc}
$\mu$ & $m_f$ & $m_\pi$ & $\rho$ & $\sigma$ & $S$ & $\epsilon$\\
\tableline
.50 & .236(1) & .188(2) & .00045(2) & .796(5) & 1.36(1) & .217(1)\\
.60 & .126(1) & .24(6) & .0069(4) & .783(4) & 1.34(1) & .223(1)\\
.65 & .060(4) & .21(1) & .0163(4) & .765(4) & 1.31(1) & .232(1)\\
.70	& 0 & .21(1) & .0905(4) & .586(4) & 1.10(2) & .303(1)\\
.725 & \ & .40(2) & .230(1) & .131(4) & .79(2) & .426(1)\\
.75 & \ & .49(4) & .267(1) & .068(3) & .77(1) & .452(1)\\
.775 & \ & .55(8) & .299(2) & .050(3) & .75(1) & .475(1)\\
.80 & \ &.78(5) & .337(2) & .038(3) & .74(1) & .501(2)\\
.825 & \ & .78(5) & .374(2) & .026(2) & .74(1) & .526(2)\\
.85 & \ & --- & .416(2) & .020(2) & .73(1) & .556(2)
\end{tabular}
\end{table}

\begin{table}
\caption{\
Same as Table II except $1/g^2=0.80$.}
\label{3.}
\begin{tabular}{dcccccc}
$\mu$ & $m_f$ & $m_\pi$ & $\rho$ & $\sigma$ & $S$ & $\epsilon$\\
\tableline
.15 & .137(2) & .22(1) & .0026(2) & .287(1) & .532(1) & .303(1)\\
.20 & .075(3) & .20(1) & .0047(1) & .274(1) & .525(1) & .305(1)\\
.225 & .041(4) & .26(2) & .0078(1) & .262(2) & .518(1) &.308(1)\\
.25 & .010(5) & .22(1) & .0120(7) & .245(1) & .511(1) & .310(1)\\
.275 & 0 & .24(1) & .0162(6) & .220(1) & .501(1) & .314(1)\\
.30 & \ & .25(2) & .0220(3) & .190(1) & .490(1) & .319(1)\\
.325 & \ & .28(3) & .0288(2) & .160(1) & .480(1) & .323(1)\\
.35 & \ & .30(3) & .0375(4) & .129(1) & .472(1) & .329(1)\\
.40 & \ & .35(3) & .056(1) & .083(1) & .461(1) & .339(1)\\
.50 & \ &.65(5) & .099(1) & .042(1) & .452(1) & .360(1)
\end{tabular}
\end{table}

\begin{table}
\caption{\
Same as Table I except $N_{f{\rm latt}}= 1$.}
\label{4.}
\begin{tabular}{dccccc}
$1/g^2$ & $m_f$ & $m_\pi$ & $\sigma$ & $S$ & $\epsilon$\\
\tableline
1.00 & .070(1) & .25(5) & .064(2) & 1.063(1) & .314(1)\\
.95 & .075(5) & .22(3) & .078(2) & 1.121(1) & .313(1)\\
.90 & .105(2) &.23(3) & .106(5) & 1.193(1) & .309(1)\\
.85 & .14(2) & .20(2) & .141(5) & 1.277(1) & .305(1)\\
.80 & .192(4) & .21(2) & .200(7) & 1.381(2) & .298(2)\\
.70 & .32(1) & .15(2) & .336(9) & 1.655(2) & .280(2)\\
.60 & .475(5) & .26(5) & .511(12) & 2.065(2) & .253(3)\\
.50 & .650(5) & .15(1) & .706(15) & 2.675(2) & .220(3)
\end{tabular}
\end{table}

\begin{table}
\caption{\
Same as Table II except $N_{f{\rm latt}}=1$ and $1/g^2=0.60$.}
\label{5.}
\begin{tabular}{dcccccc}
$\mu$ & $m_f$ & $m_\pi$ & $\sigma$ & $\rho$ & $S$ & $\epsilon$\\
\tableline
.30 & .173(1) & .15(2) & .498(2) &  .00133(57) & 2.062(1) & .255(1)\\
.35 & .116(2) &.20(5) & .491(2) & .00432(19) & 2.055(1) & .257(1)\\
.40 & .046(2) & .15(2) & .470(2) & .0104(3) & 2.038(1) & .252(1)\\
.425 & 0 & --- & .439(4) & .0171(4) & 2.018(2) & .268(1)\\
.45 & \ & .19(2) & .373(15) & .0312(3) & 1.977(4) & .281(1)\\
.475 & \ & .22(2) & .259(15) & .0541(8) & 1.924(3) & .300(1)\\
.4875 & \ & .25(3) & .206(15) & .0644(3) & 1.904(2) & .308(1)\\
.50 & \ & .30(2) & .178(5) & .0731(5) & 1.890(2) & .313(1)\\
.525 & \ & .30(2) & .124(3) & .0892(4) & 1.871(1) & .325(1)\\
.55 & \ & .36(3) & .091(5) & .1034(5) & 1.859(1) & .333(1)\\
.60 &\ & .52(4) & .060(7) & .1310(7) & 1.843(1) & .350(1)
\end{tabular}
\end{table}

\begin{table}
\caption{\
Same as Table V except $1/g^2=0.070$.}
\label{6.}
\begin{tabular}{dcccccc}
$\mu$ & $m_f$ & $m_\pi$ & $\sigma$ & $\rho$ & $S$ & $\epsilon$\\
\tableline
.15 & .164(1) & .22(1) & .331(1) & .00067(33) & 1.65(1) & .281(1)\\
.20 & .113(5) & .22(1) & .324(1) & .00247(64) & 1.65(1) & .282(1)\\
.225 & .085(5) & .21(1) & .315(1) & .00468(9) & 1.64(1) & .283(1)\\
.25 & .040(8) & .22(2) & .301(1) & .00689(43) & 1.63(1) & .284(1)\\
.275 & --- & .24(1) & .273(3) & .0102(3) & 1.62(1) & .289(1)\\
.30 & 0 & .26(2) & .235(7) & .0175(4) & 1.61(1) & .294(1)\\
.325 & \ & .30(1) & .195(5) & .0235(7) & 1.60(1) & .299(1)\\
.35 & \ & .25(2) & .162(7) & .0309(6) & 1.59(1) & .304(1)\\
.40 & \ & .38(1) & .107(7) & .0490(3) & 1.58(1) & .314(1)\\
.50 & \ & .50(5) & .052(3) & .0898(3) & 1.56(1) & .335(1)
\end{tabular}
\end{table}\eject
\centerline{\bf Figure Captions}

\noindent
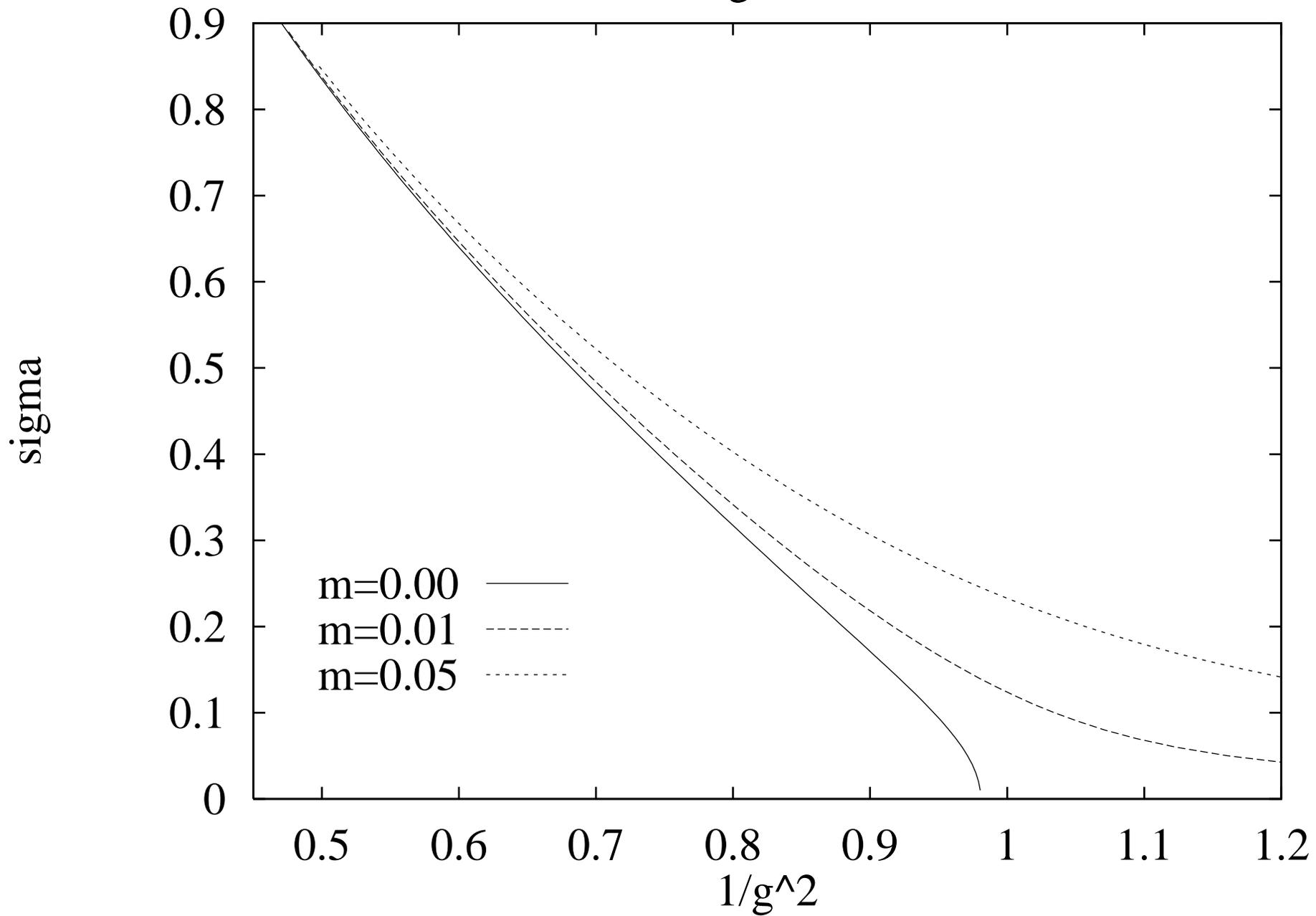
Figure 1: Plot of $\langle\sigma\rangle$ versus $1/g^2$ at $\mu=0$
evaluated using
the gap equation (2.14) on a $16^3$ lattice for three different bare fermion
masses $m$.

\noindent
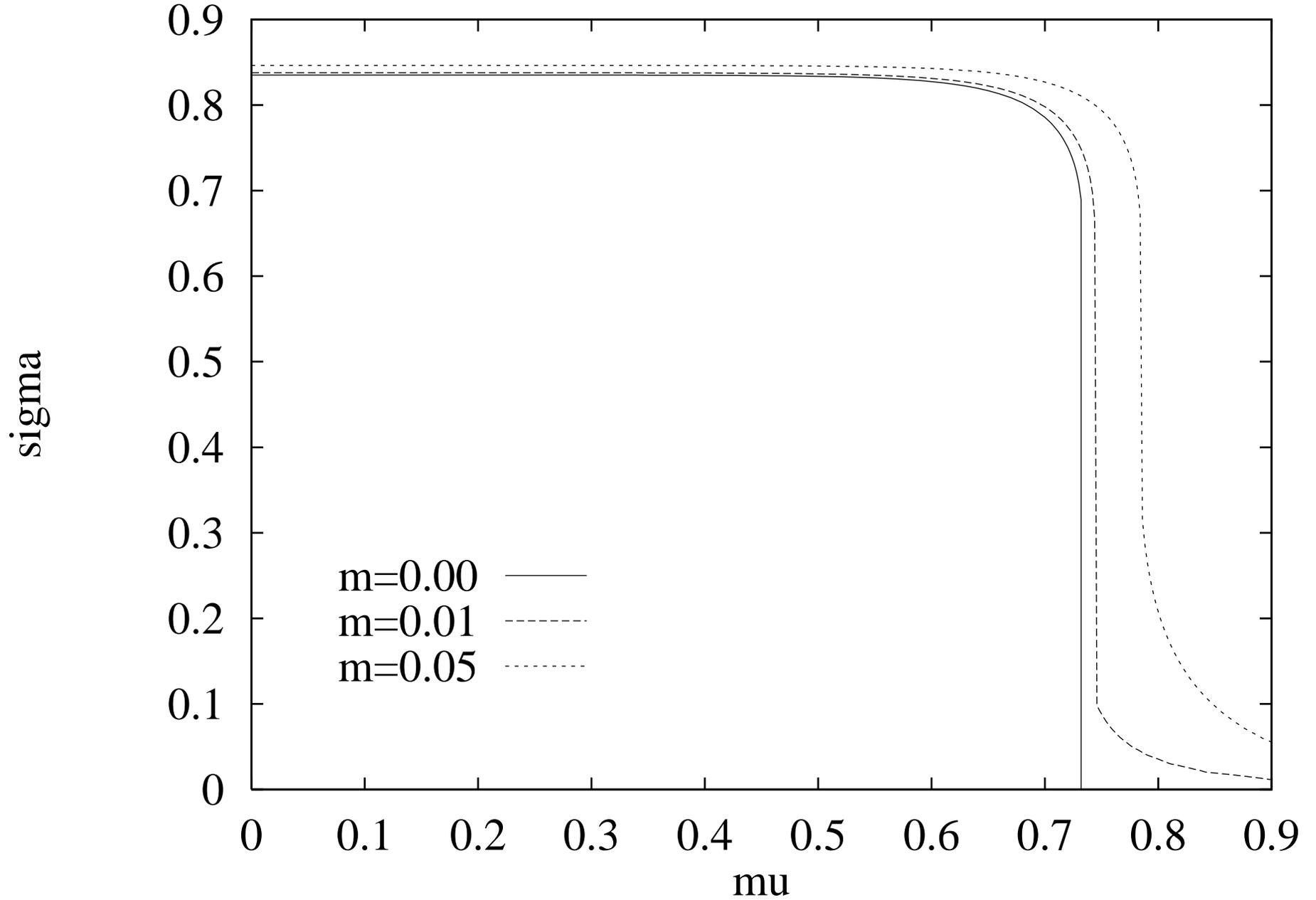
Figure 2: Gap equation prediction of
$\langle\sigma\rangle$ versus $\mu$ at $1/g^2=0.5$
for three different $m$.

\noindent
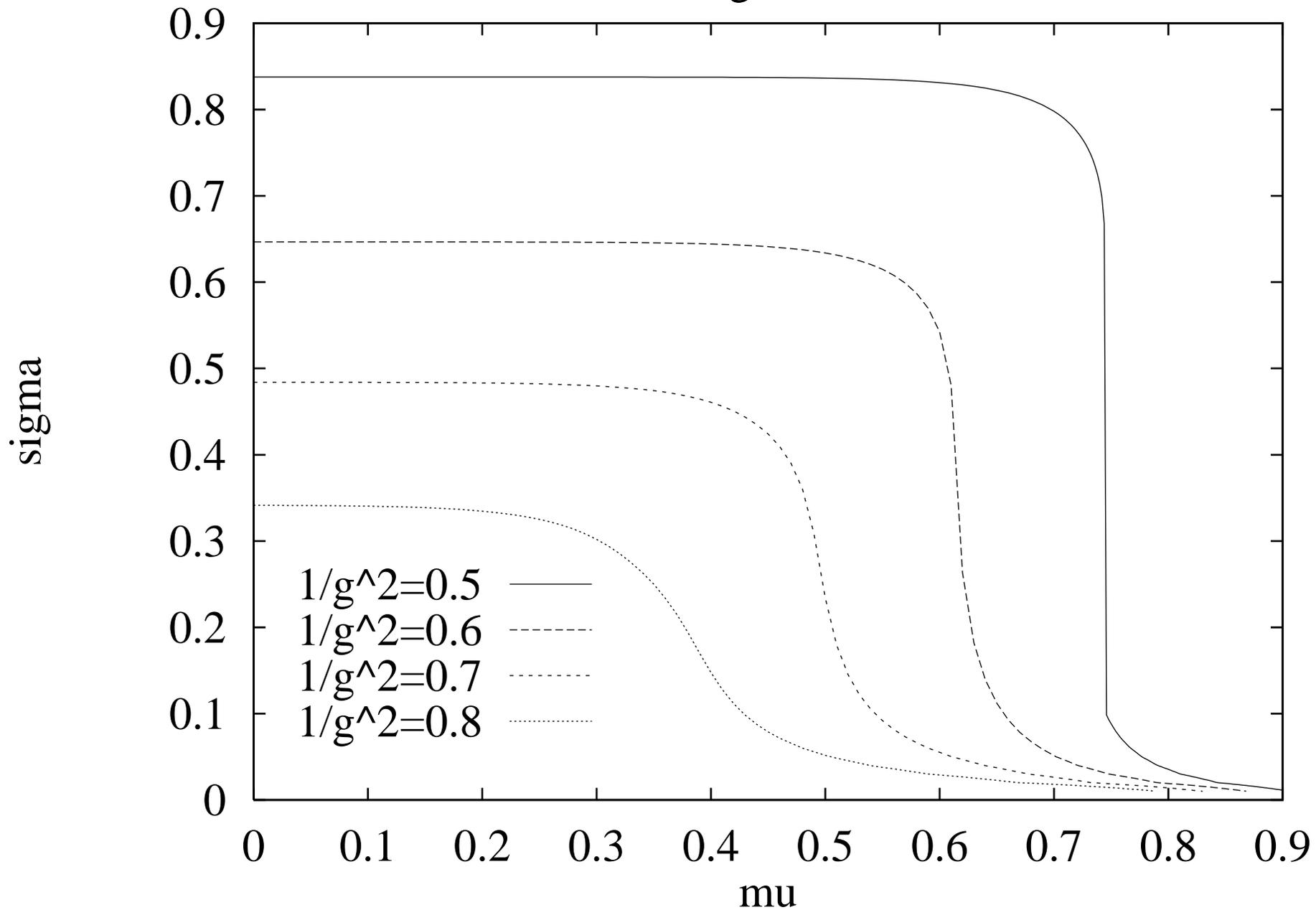
Figure 3: Gap equation prediction of
$\langle\sigma\rangle$ versus $\mu$ for $m=0.01$
at four different
$1/g^2$.

\noindent
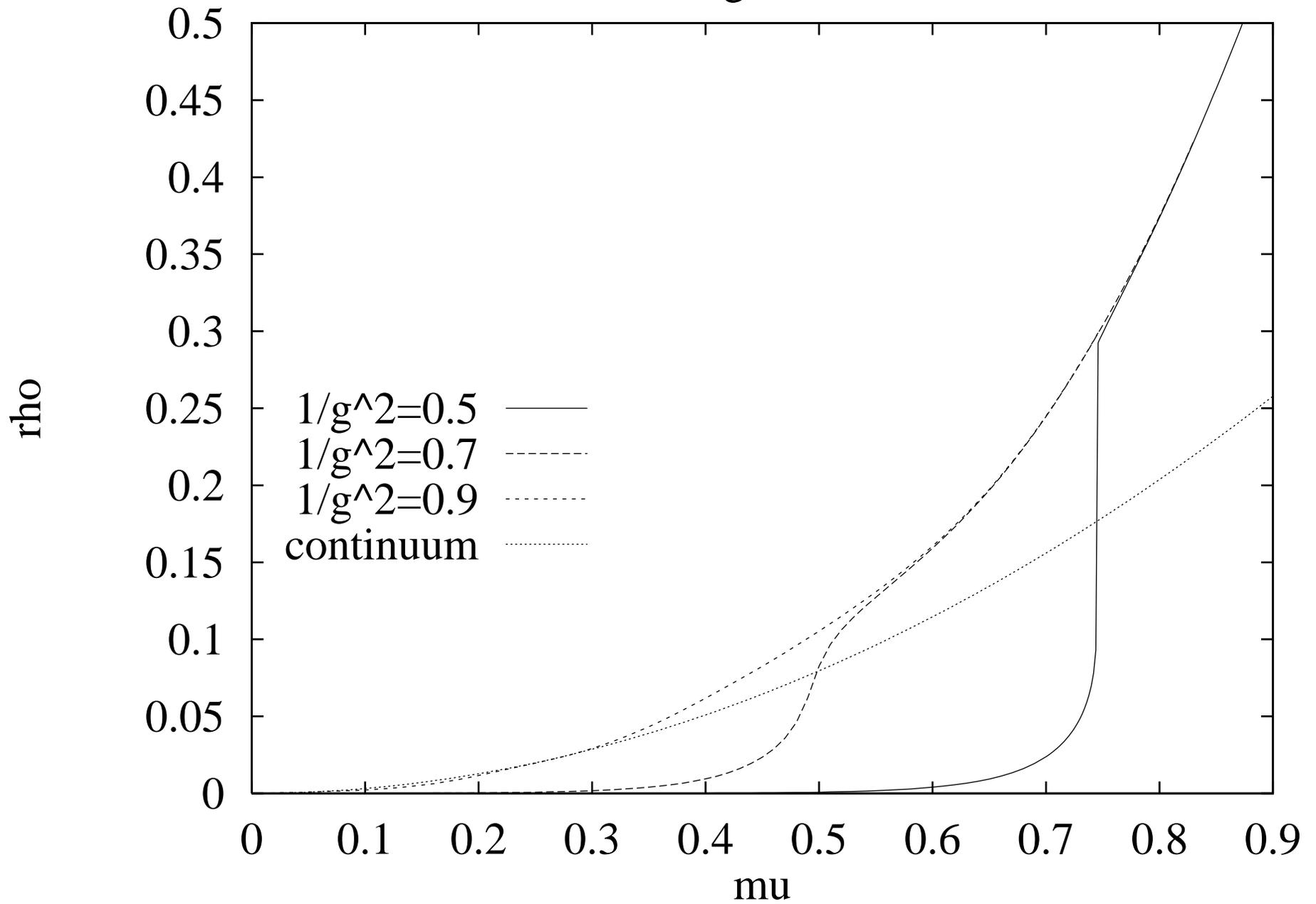
Figure 4: Gap equation prediction of number density $\rho$
versus $\mu$  at three different
$1/g^2$, plus the continuum free-field prediction.

\noindent
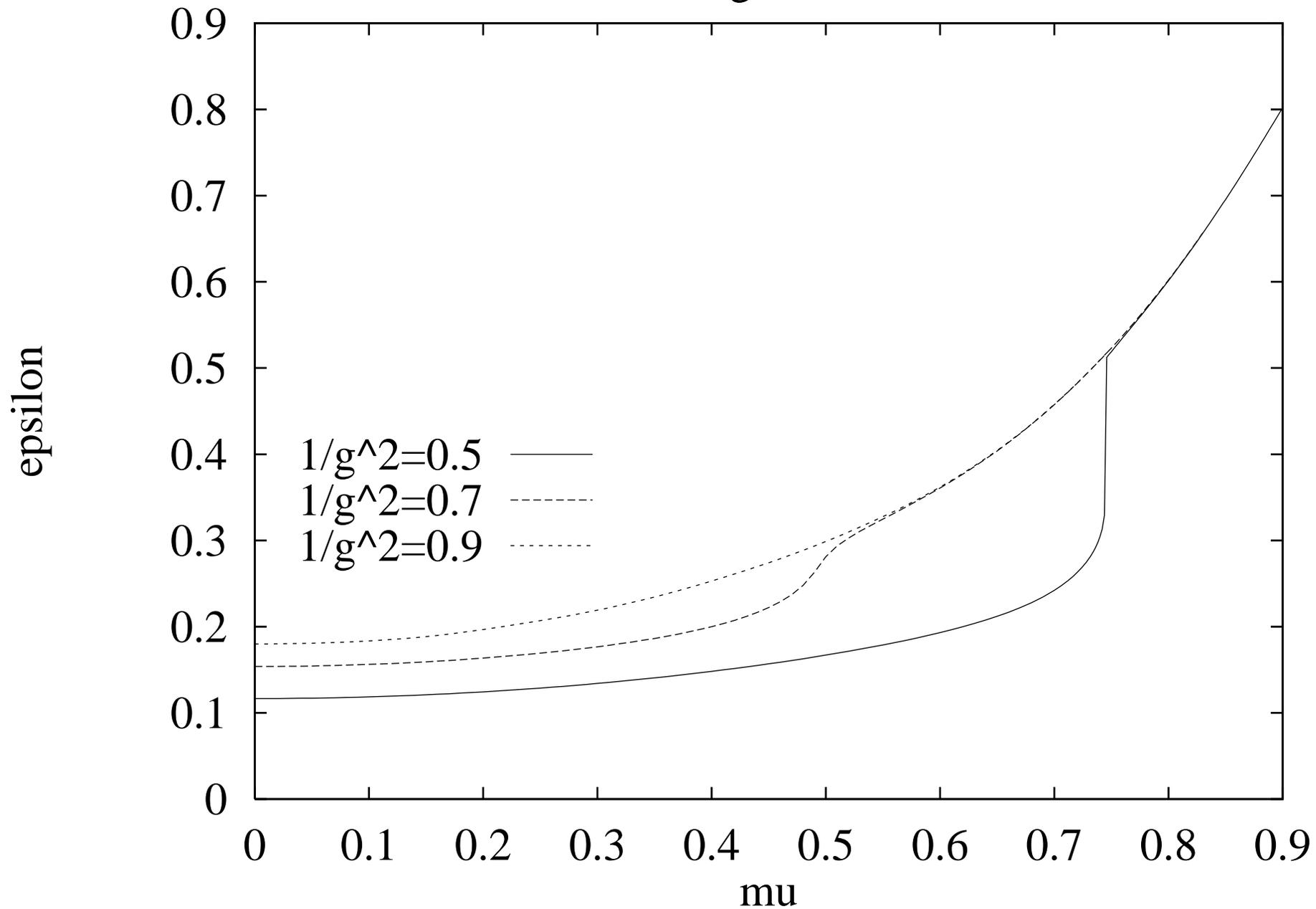
Figure 5: Gap equation prediction of energy density $\epsilon$ versus
$\mu$ at three different $1/g^2$.

\noindent
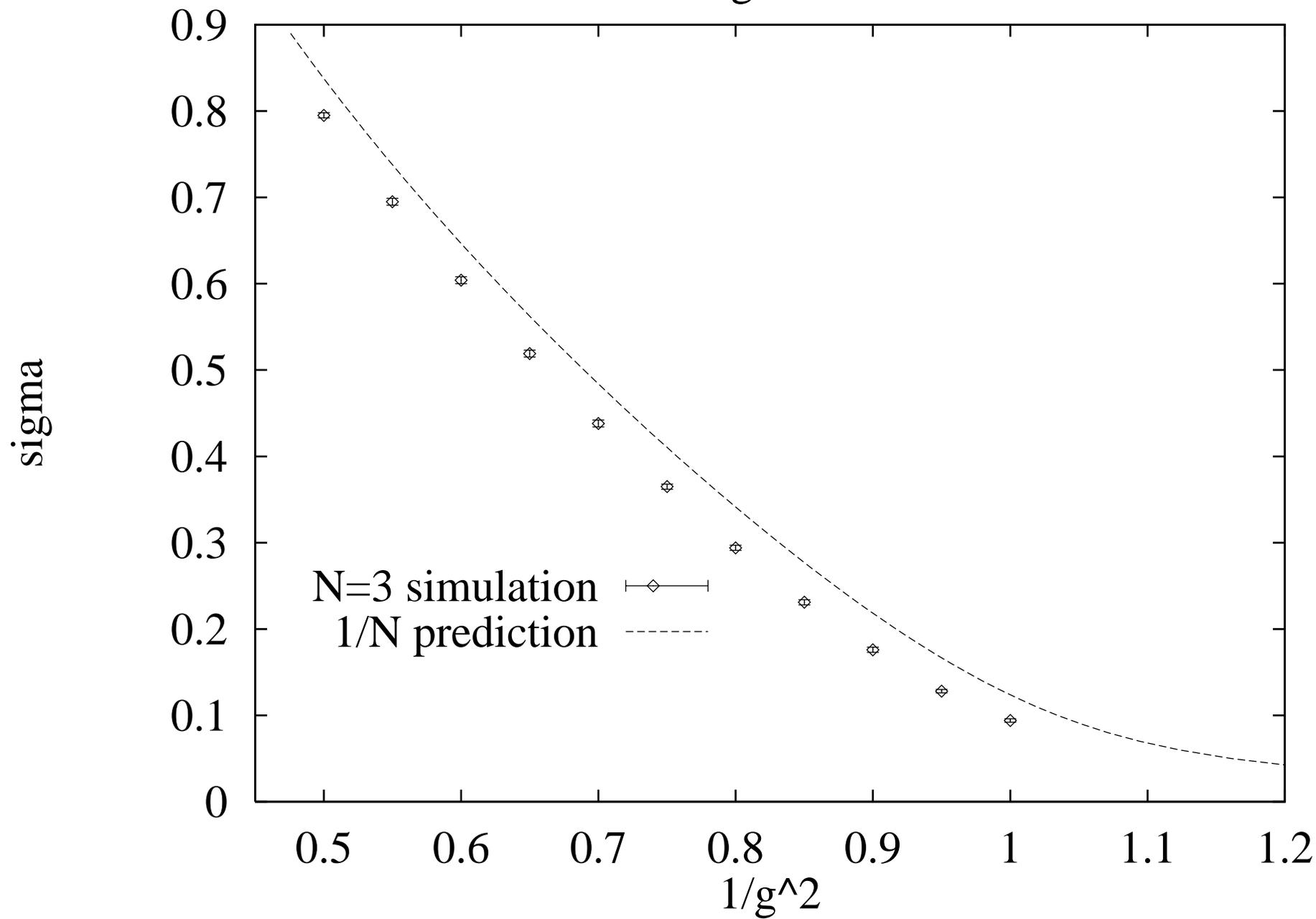
Figure 6: Simulation results for $\sigma$ versus $1/g^2$ at $\mu=0$ using
$N_{f{\rm latt}}=3$ and $m=0.01$. Also shown is the gap equation prediction
of Fig. 1.

\noindent
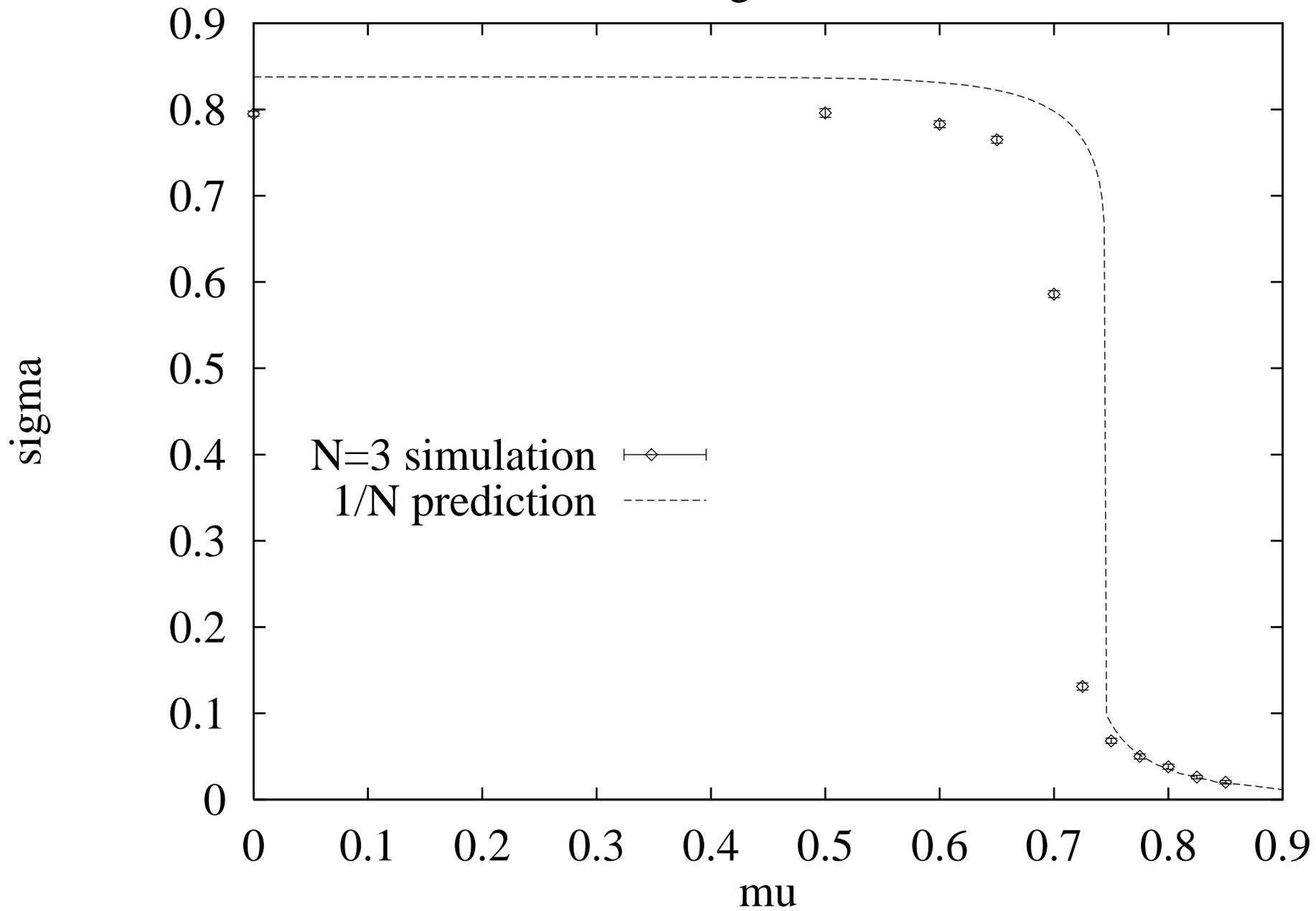
Figure 7: Simulation results for $\sigma$ versus $\mu$ at $1/g^2=0.5$ using
$N_{f{\rm latt}}=3$. Also shown is the gap equation prediction
of Fig. 3.

\noindent
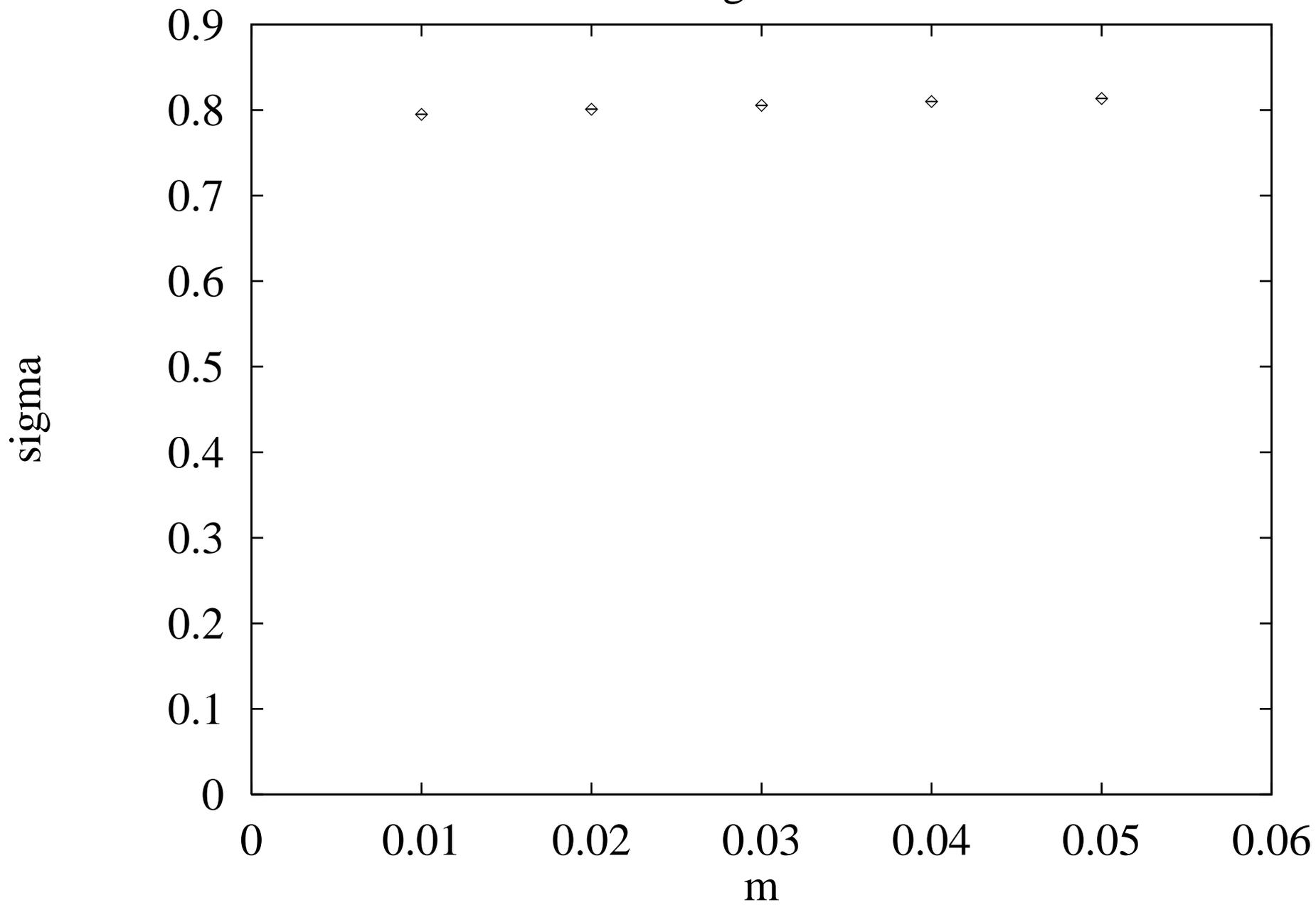
Figure 8: Simulation results for $\sigma$ versus $m$
at $1/g^2=0.5$ and $\mu=0.5$
using $N_{f{\rm latt}}=3$.

\noindent
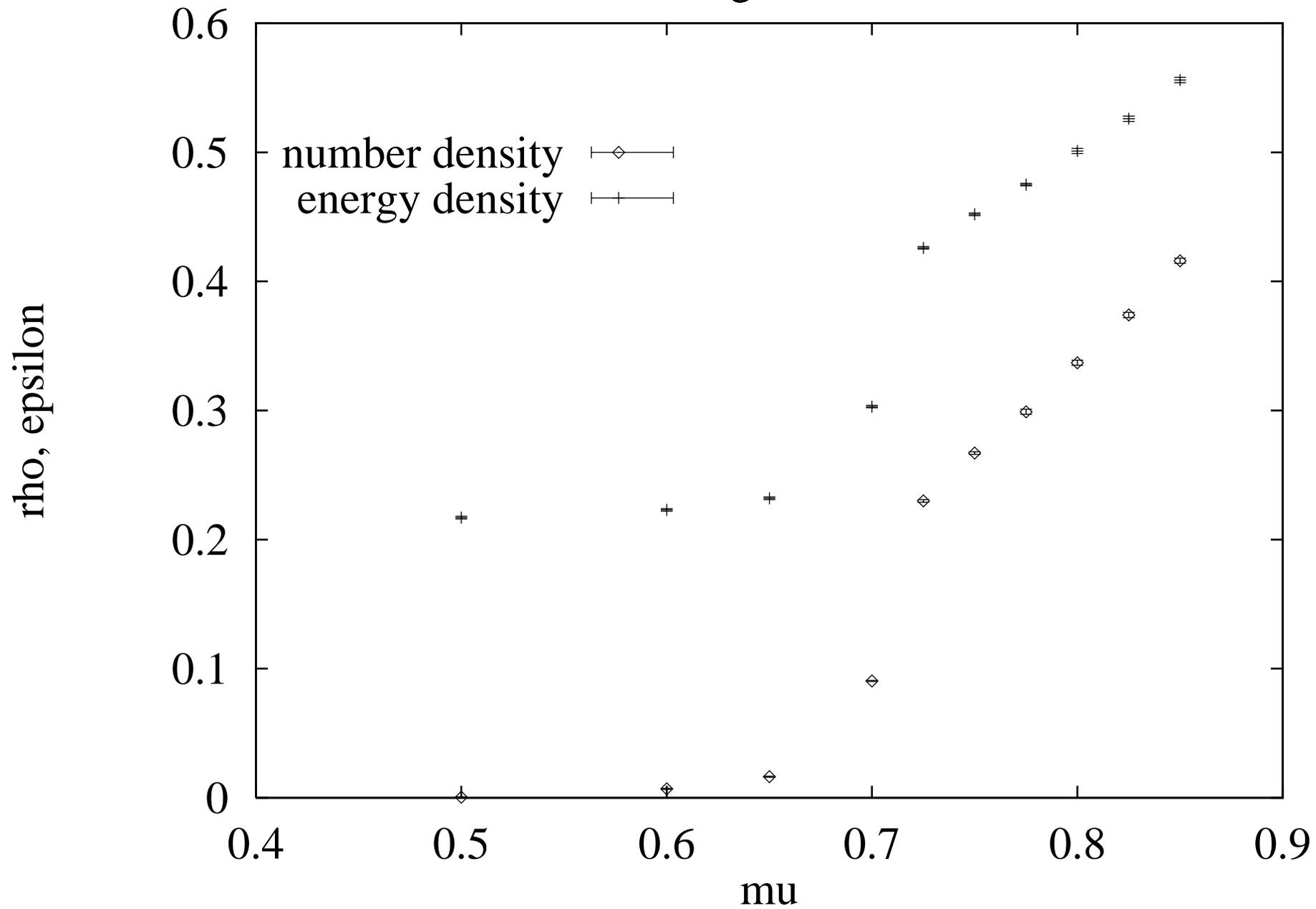
Figure 9: Simulation results for $\rho$ and $\epsilon$ versus $\mu$ at
$1/g^2=0.5$ using $N_{f{\rm latt}}=3$.

\noindent
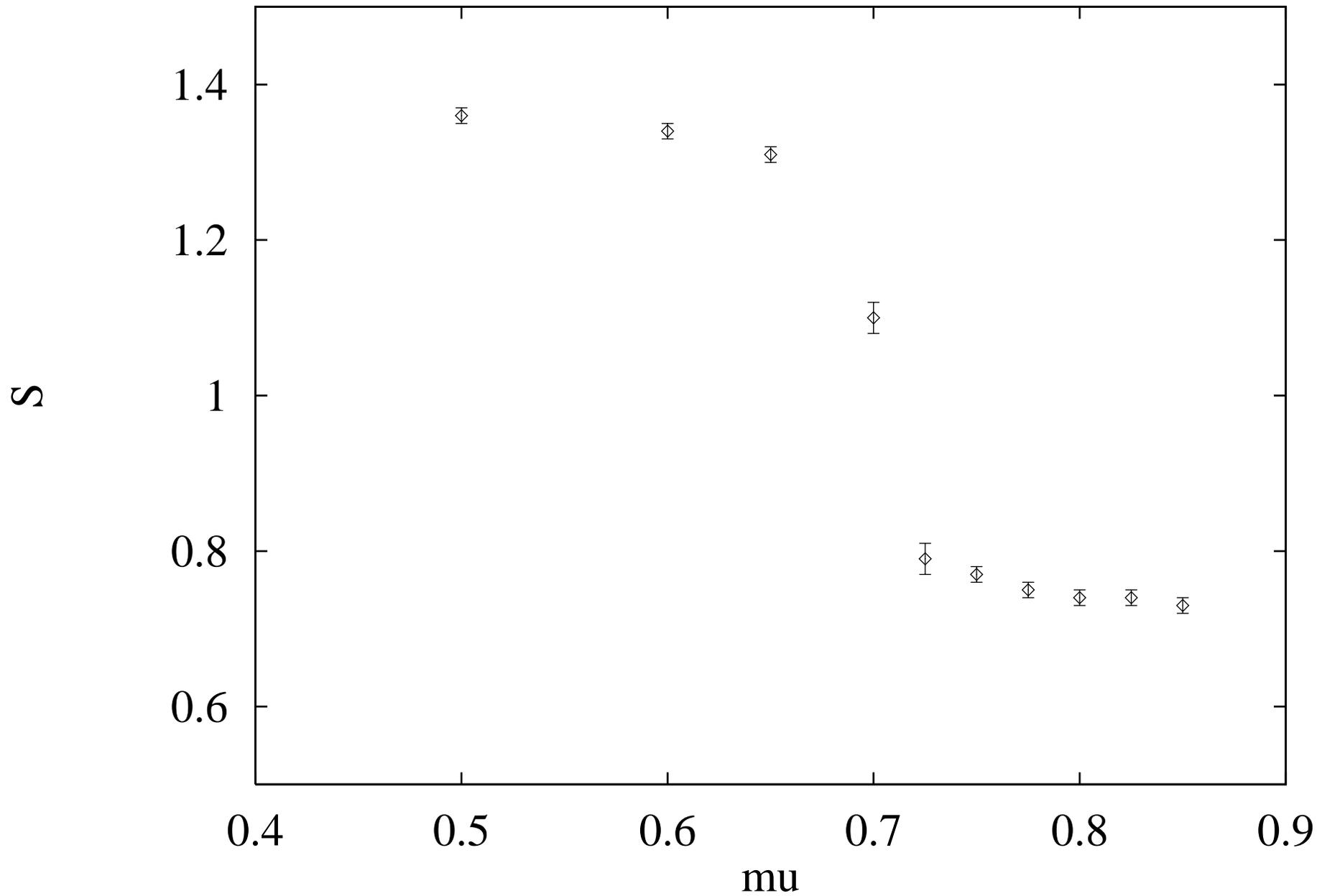
Figure 10: Simulation results for action $S$ versus $\mu$ at $1/g^2=0.5$
using $N_{f{\rm latt}}=3$.

\noindent
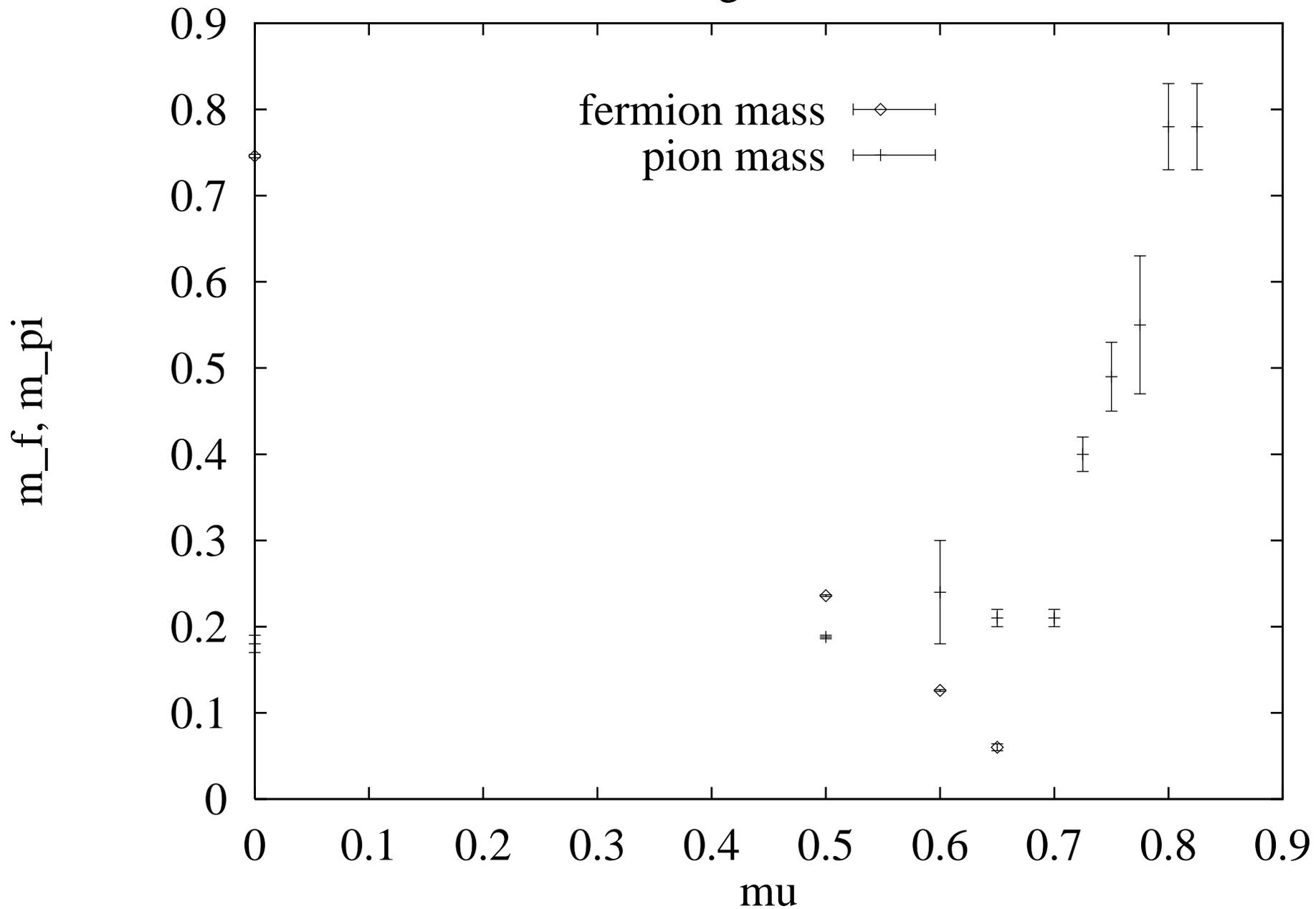
Figure 11: Simulation results for fermion mass $m_f$ and pion mass
$m_\pi$ at $1/g^2=0.5$ using $N_{f{\rm latt}}=3$.

\noindent
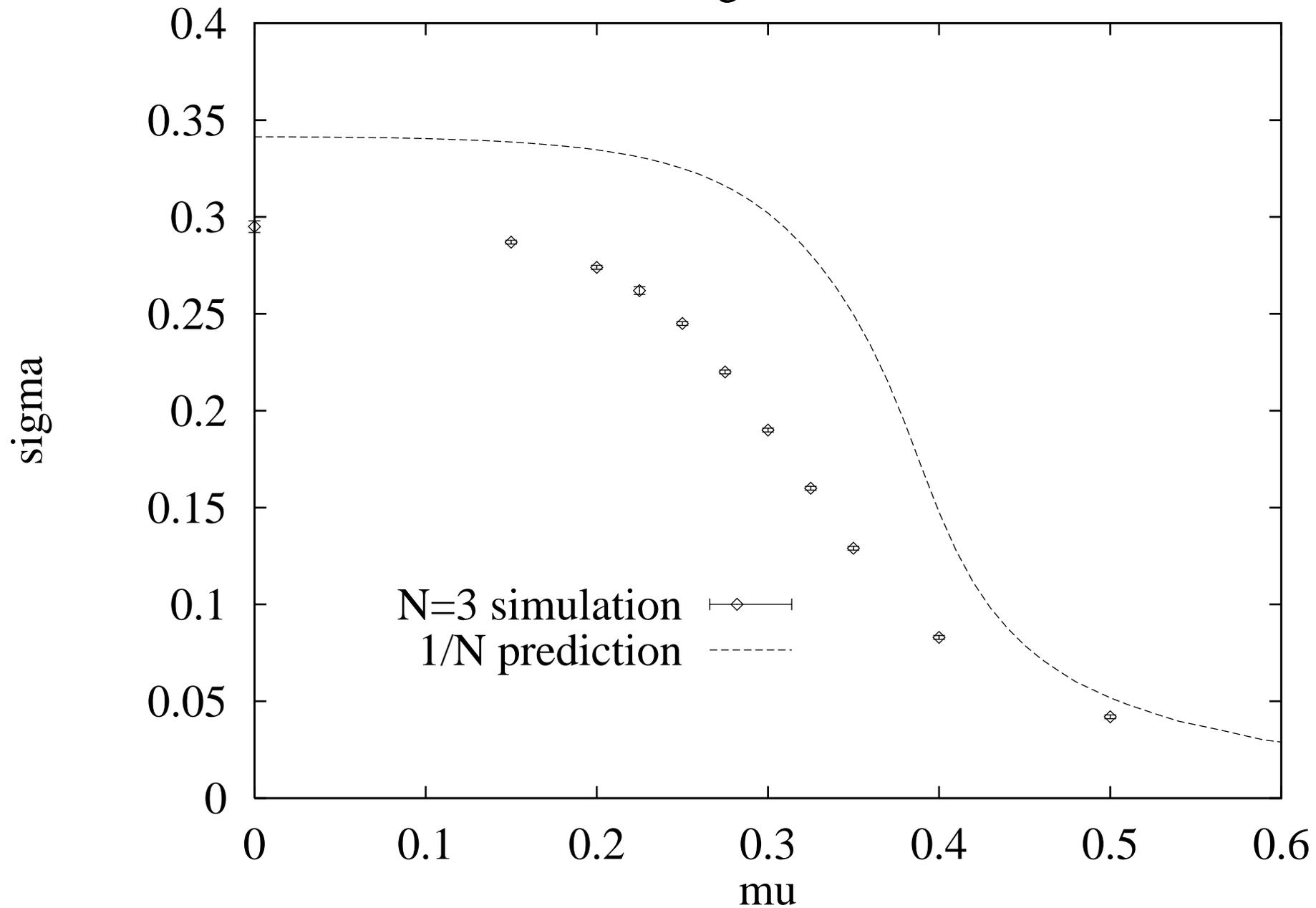
Figure 12: Same as Fig. 7 except $1/g^2=0.8$.

\noindent
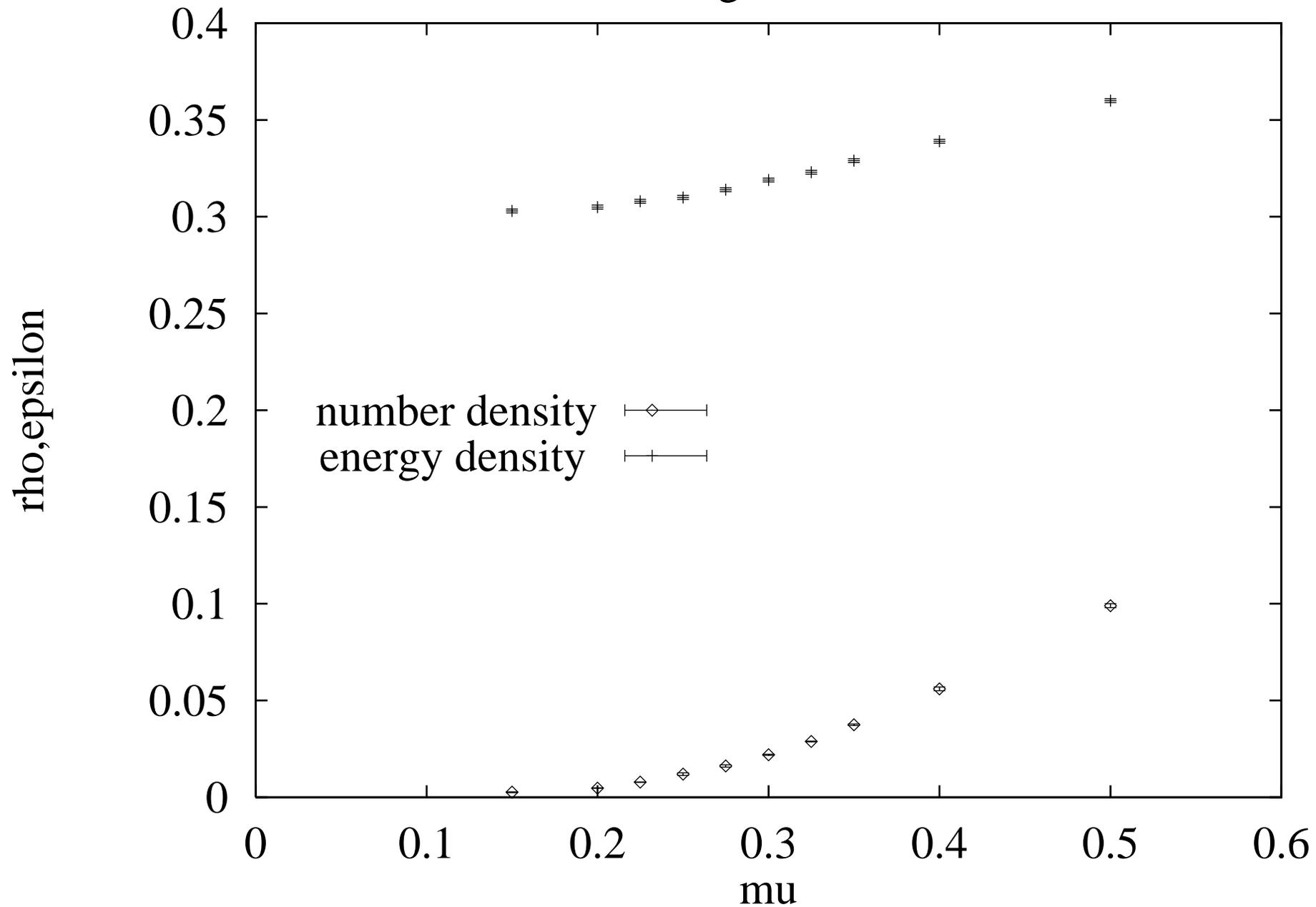
Figure 13: Same as Fig. 9 except $1/g^2=0.8$.

\noindent
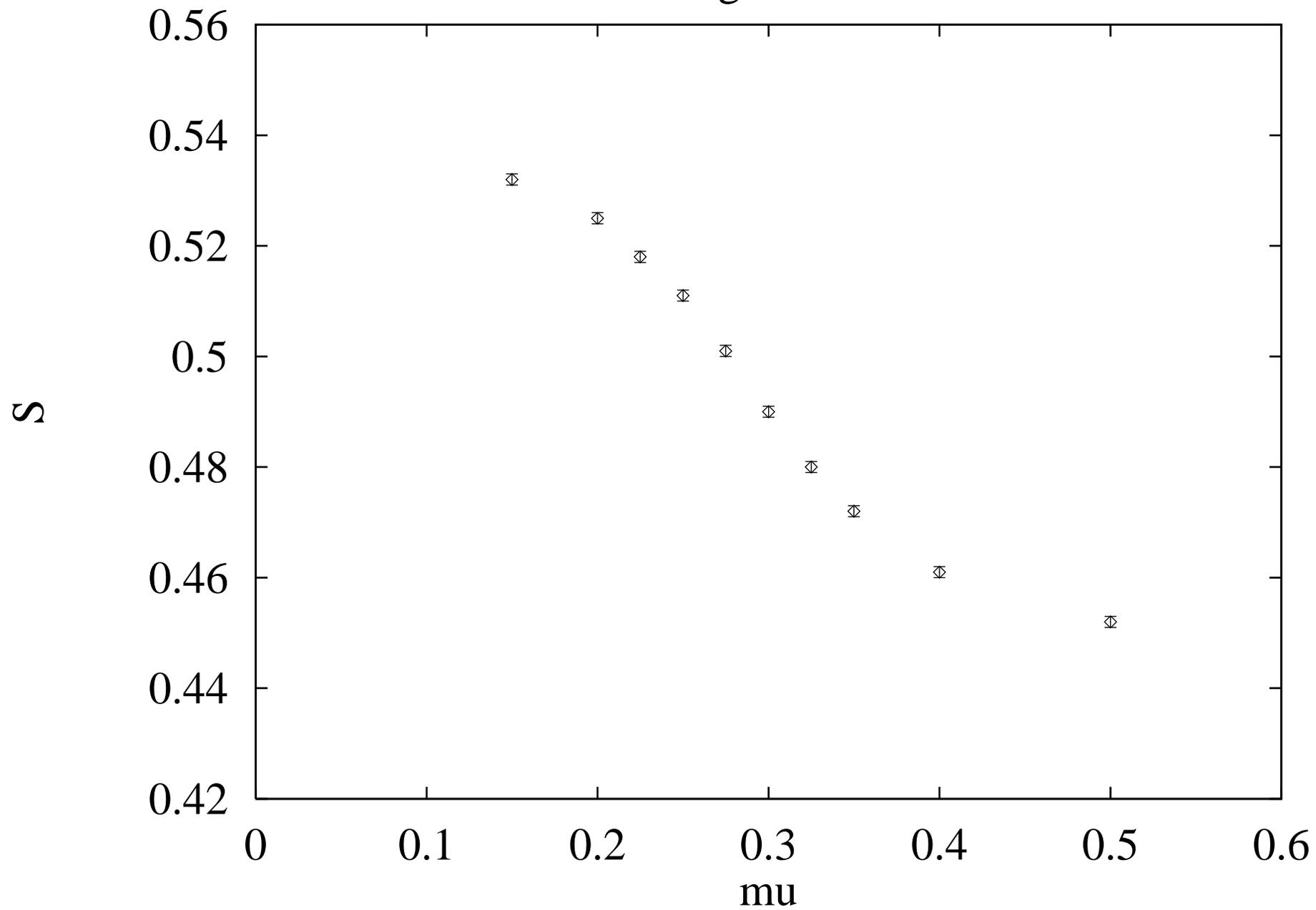
Figure 14: Same as Fig. 10 except $1/g^2=0.8$.

\noindent
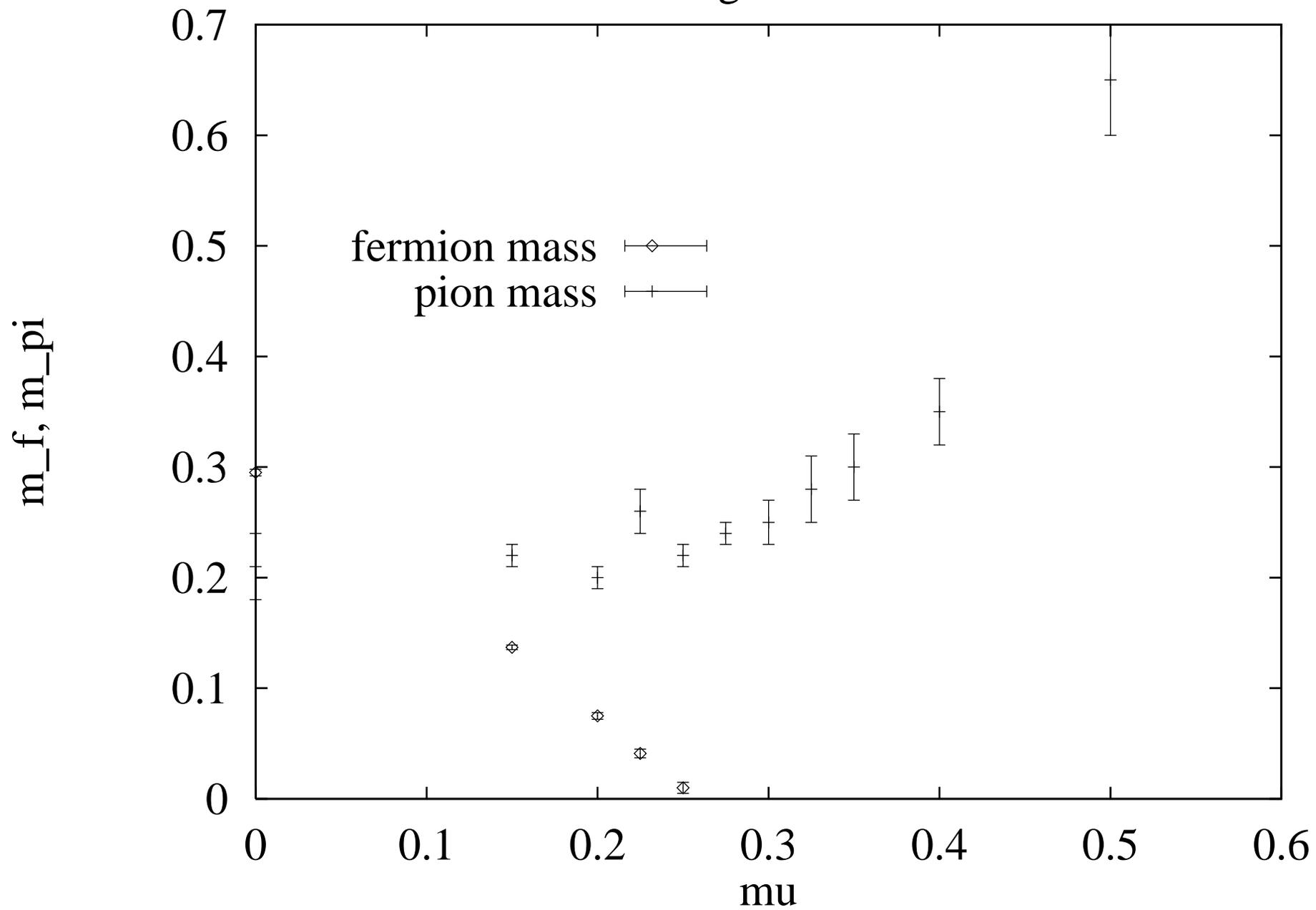
Figure 15: Same as Fig. 11 except $1/g^2=0.8$.

\noindent
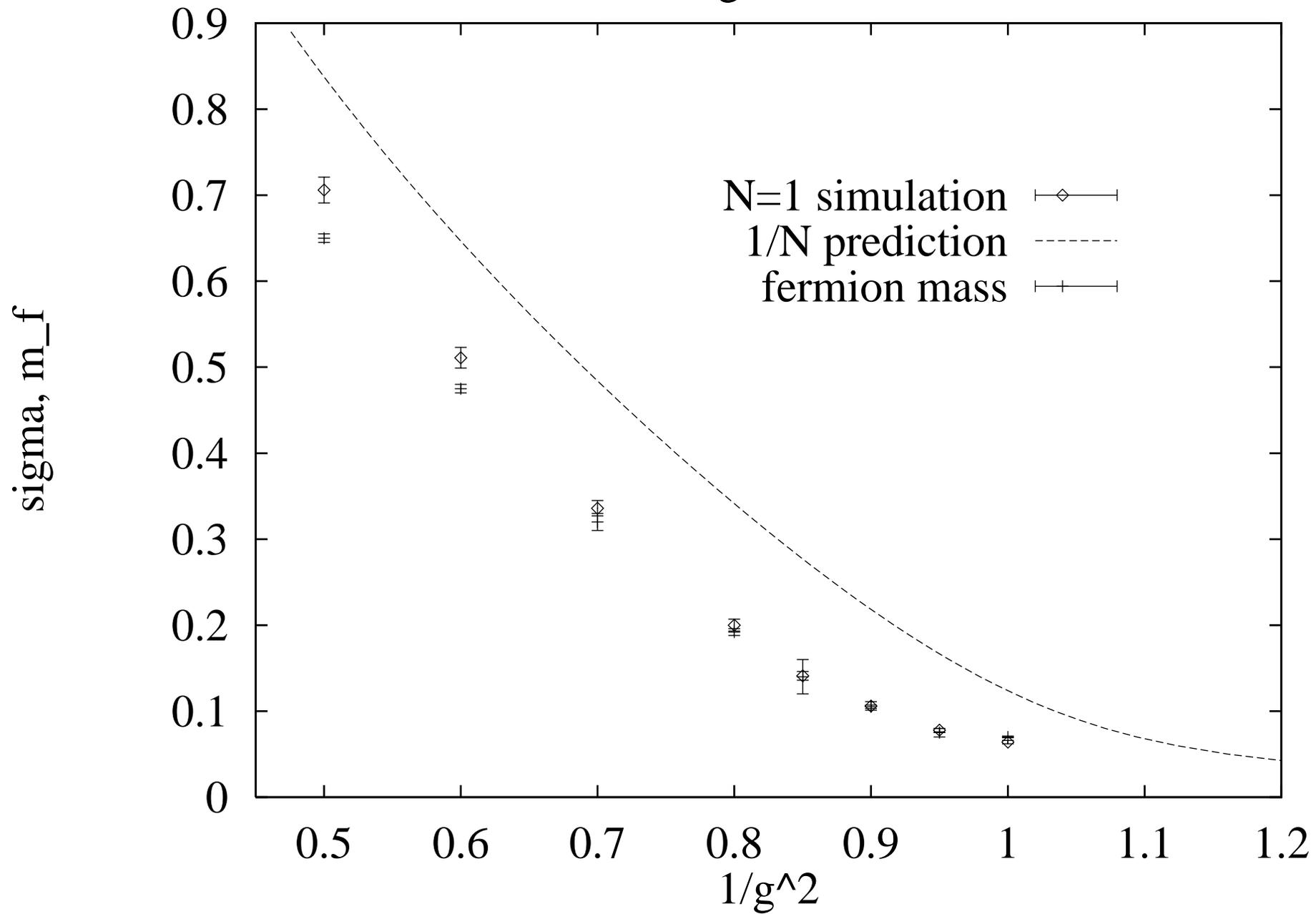
Figure 16: Same as Fig. 6 except $N_{f{\rm latt}}=1$. Also shown is
fermion mass $m_f$.

\noindent
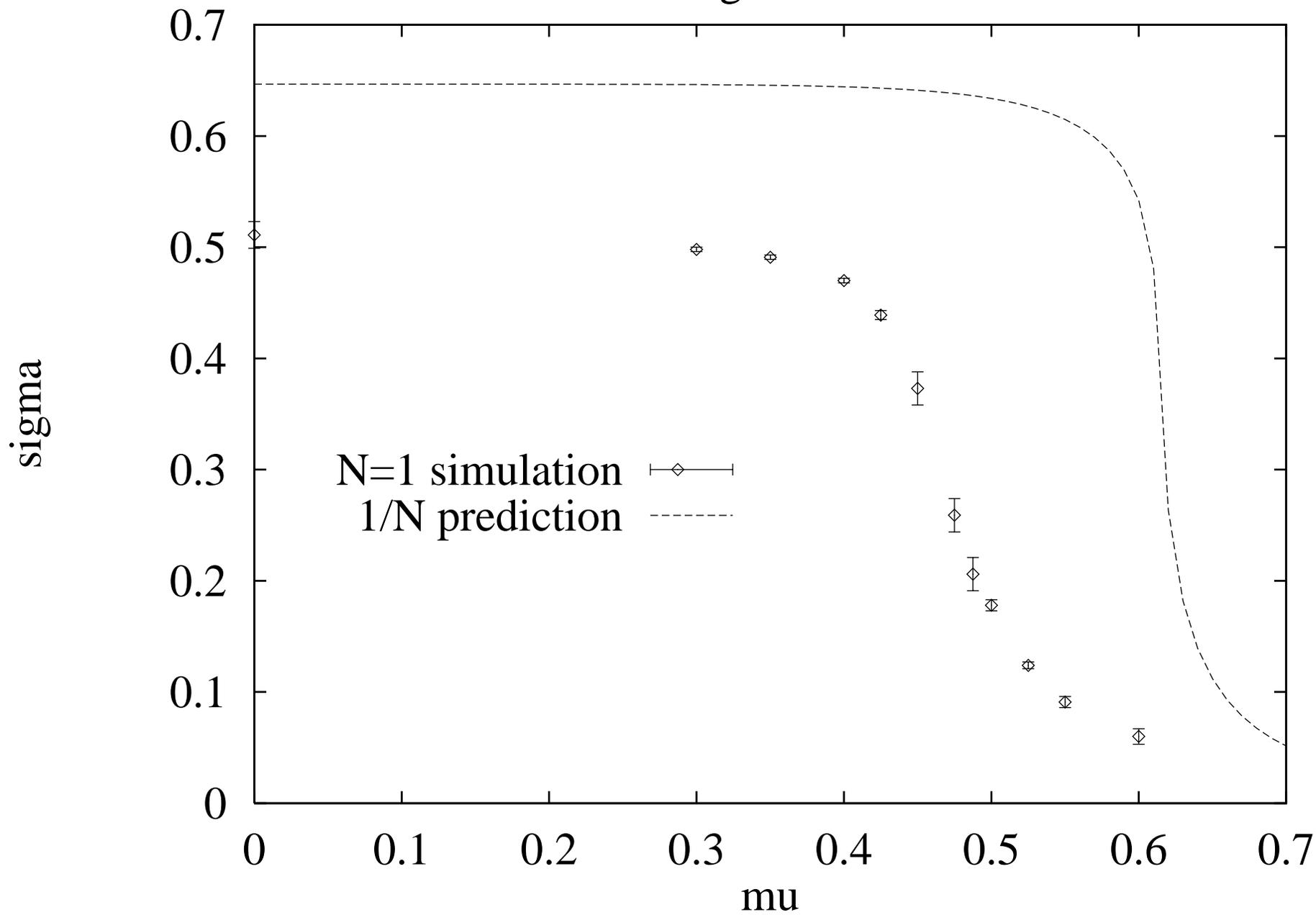
Figure 17: Same as Fig. 7 except $N_{f{\rm latt}}=1$ and $1/g^2=0.6$.

\noindent
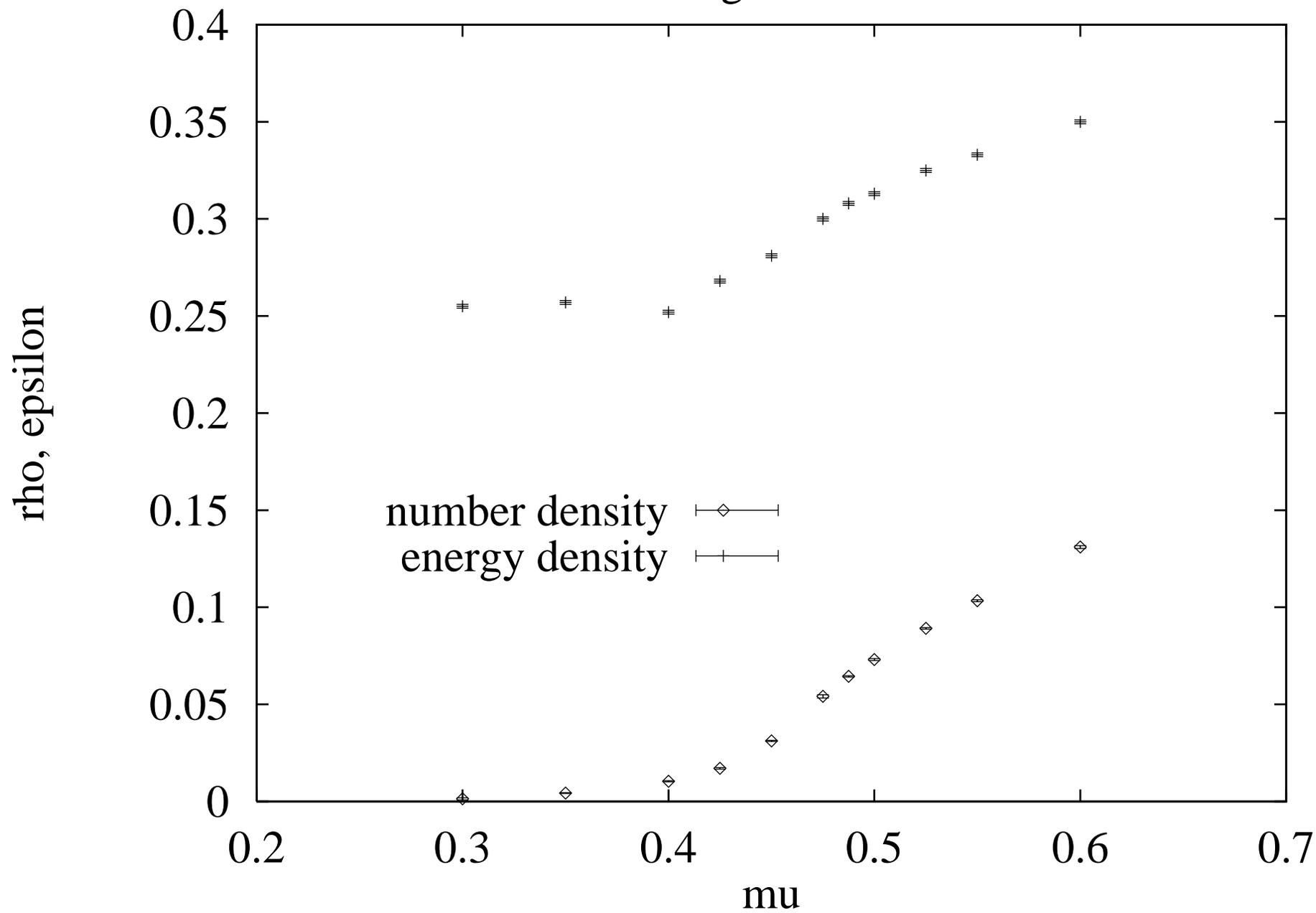
Figure 18: Same as Fig. 9 except $N_{f{\rm latt}}=1$ and $1/g^2=0.6$.

\noindent
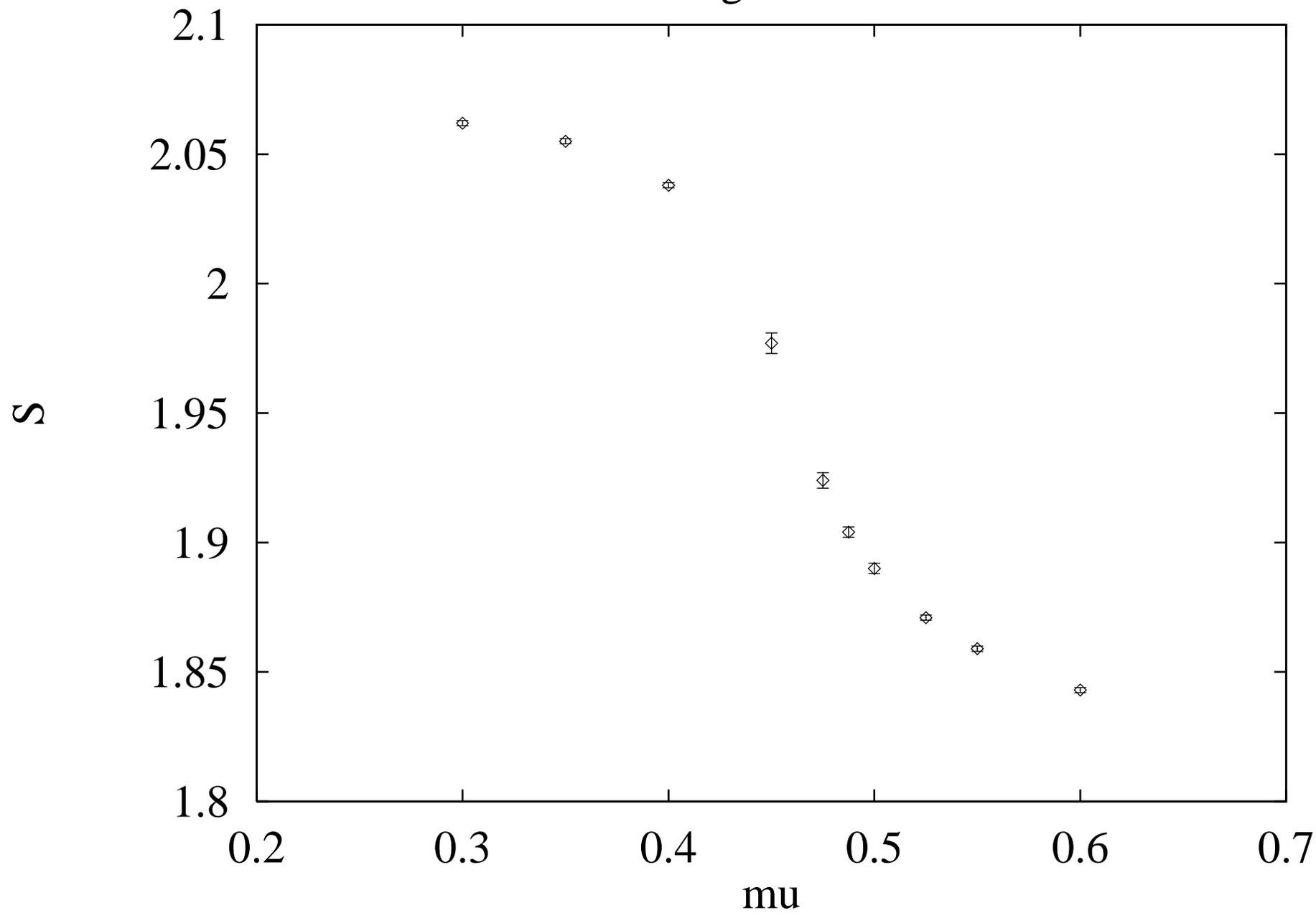
Figure 19: Same as Fig. 10 except $N_{f{\rm latt}}=1$ and $1/g^2=0.6$.

\noindent
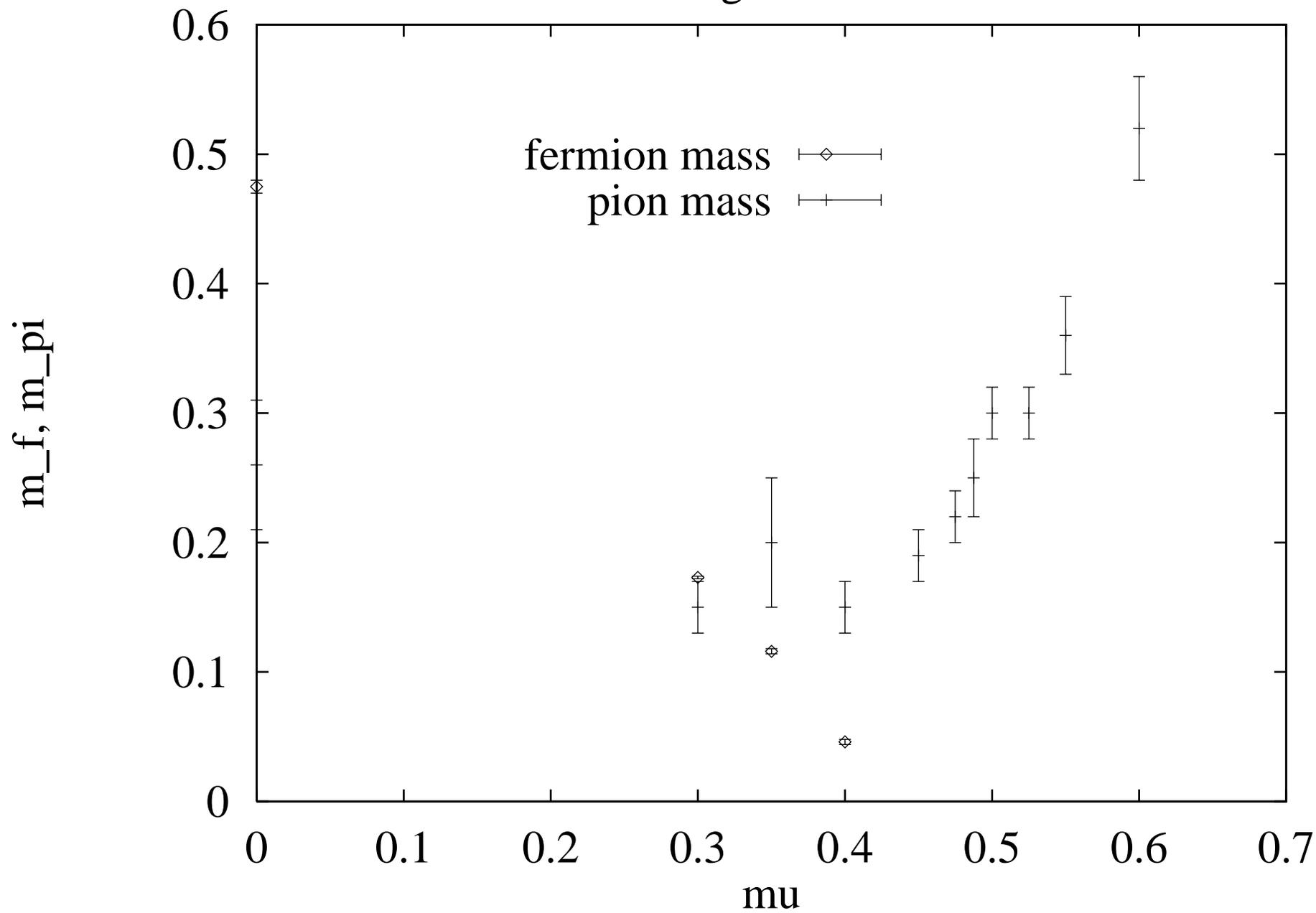
Figure 20: Same as Fig. 11 except $N_{f{\rm latt}}=1$ and $1/g^2=0.6$.

\noindent
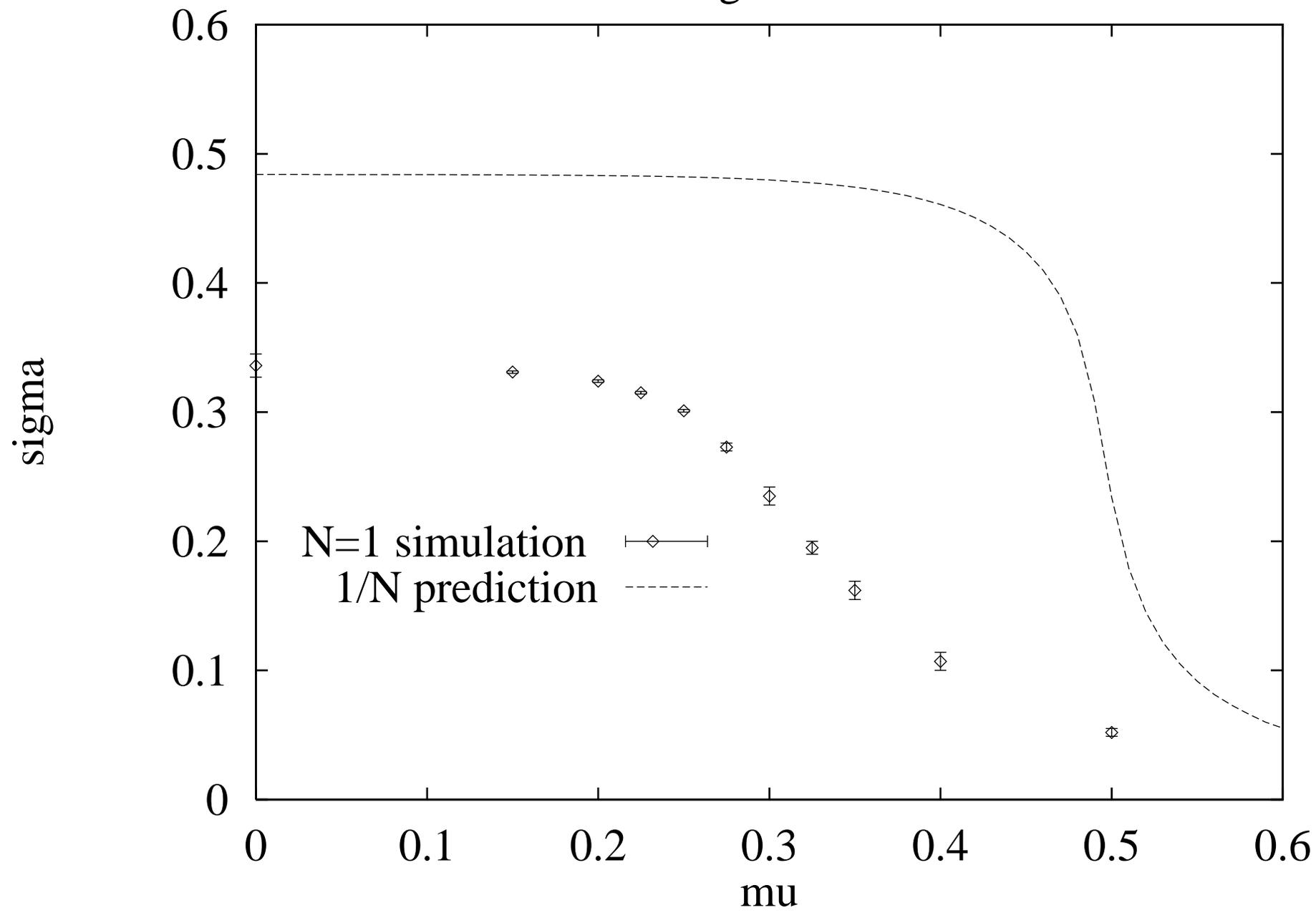
Figure 21: Same as Fig. 17 except $1/g^2=0.7$.

\noindent
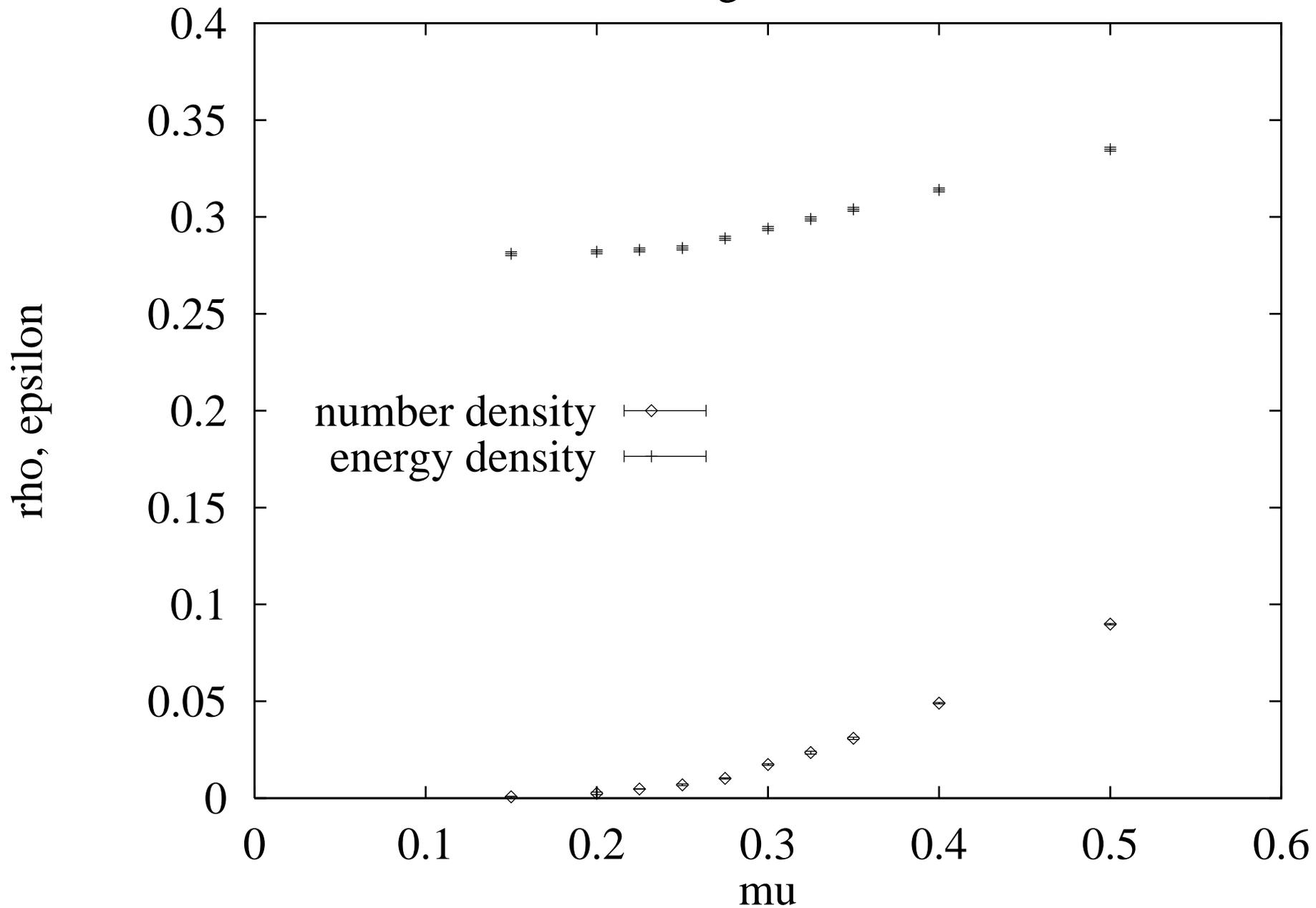
Figure 22: Same as Fig. 18 except $1/g^2=0.7$.

\noindent
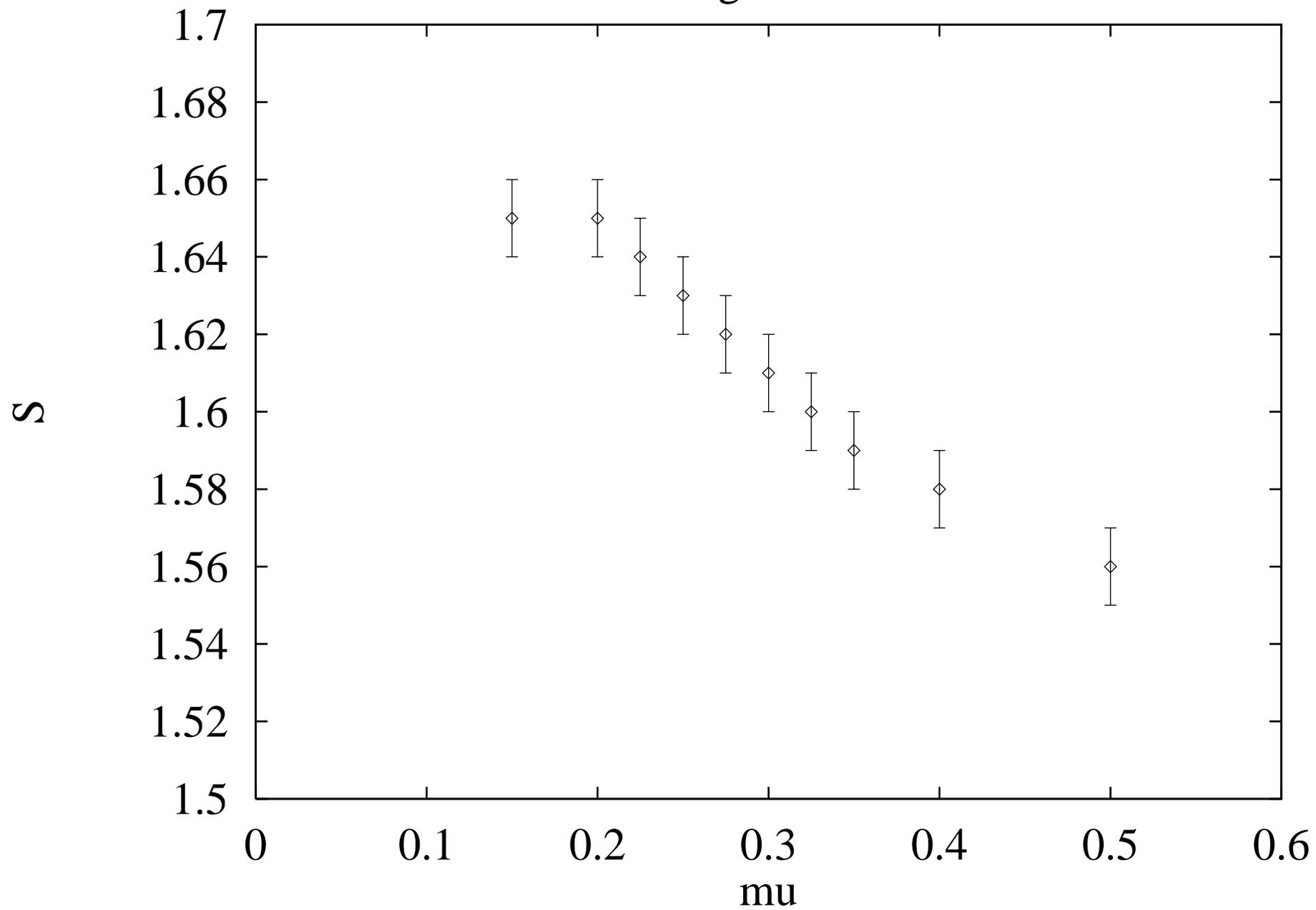
Figure 23: Same as Fig. 19 except $1/g^2=0.7$.

\noindent
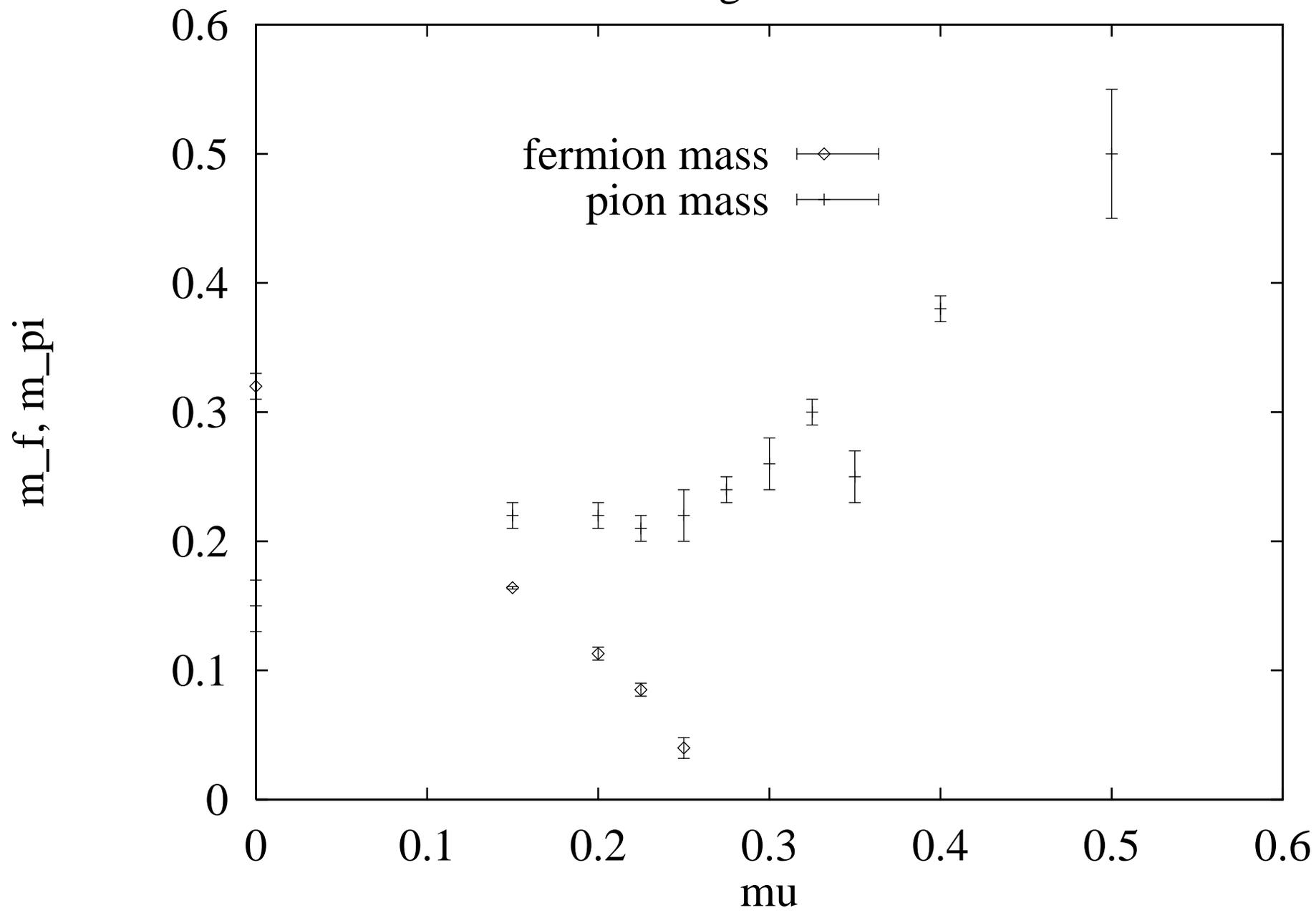
Figure 24: Same as Fig. 20 except $1/g^2=0.7$.
\end{document}